\documentclass{ws-rv9x6}
\usepackage{ws-rv-van}             
\makeindex

\RequirePackage{color}

\begin{document}

\chapter{Universal scaling relations for logarithmic-correction
exponents\label{ch1}}

\author[R. Kenna]{Ralph Kenna}

\address{Applied Mathematics Research Centre,\\
Coventry University, Coventry CV1 5FB, England \\
r.kenna@coventry.ac.uk}

\begin{abstract}
By the early 1960's advances in statistical physics had established the existence of universality classes for systems with second-order phase transitions and
characterized these by critical exponents which are different to the classical ones.
There followed the discovery of (now famous) scaling relations between the power-law critical exponents describing second-order criticality.
These scaling relations are of fundamental importance and now form a cornerstone of statistical mechanics.
In certain circumstances, such scaling behaviour is modified by multiplicative logarithmic corrections.
These are also characterized by critical exponents, analogous to the standard ones.
Recently scaling relations between these logarithmic exponents have been established.
Here, the theories associated with these advances are presented and expanded and
 the status of investigations into logarithmic corrections in a variety of models is reviewed.
\end{abstract}

\body

\section{Introduction}
\label{Introduction}

Phase transitions\index{phase transition} are abundant in nature.
They are involved in the evolution of the universe and in a multitude of phenomena in the physical, biological and socio-economic sciences.
First-order\index{phase transition!second-order}, temperature-driven, phase transitions, such as melting\index{melting} and evaporation\index{evaporation}, exhibit discontinuous changes in the internal energy through emission or absorption of a latent heat as the transition point is traversed.
Higher-order transitions, in contrast, involve a continuous change in the internal energy.
Unlike first-order transitions, they can involve divergences at the transition temperature $T_c$.
One of the major achievements of statistical physics is the fundamental explanation of such phenomena and, 150 years after their experimental discovery\cite{BeHe09}, their ubiquity ensures that their study remains one of the most exciting areas of modern physics.
This review concerns such higher-order phase transitions.

Second-order transitions\index{phase transition!second-order} are particularly common and examples include ferromagnets\index{ferromagnets}, superconductors\index{superconductors} and superfluids\index{superfluids} in  three-dimensional condensed-matter physics, as well as the Higgs\index{Higgs} phenomenon in four-dimensional particle physics.
Indeed, the edifice of lattice quantum gauge theory relies upon second-order phase transitions to achieve a continuum limit.
The Kosterlitz-Thouless\index{Kosterlitz-Thouless} transitions of a Coulomb gas in two dimensions or in thin films of $^4$He are examples of infinite-order phase transitions.

A cornerstone in the study of phase transitions is the principal of universality\index{universality} .
This maintains that entire families of systems behave identically in the neighbourhood of a Curie or critical point.\index{critical point}\index{Curie}
Nearby, thermodynamic observables and critical exponents, which characterise the transition, do not depend on the details of intermolecular interactions.
Instead they depend only on the range of interactions, inherent symmetries of the Hamiltonian and  dimensionality $d$ of the system.
Universality arises as the system develops thermal or quantum fluctuations of all sizes near the critical point, which wash out the details of interaction and render the system scale invariant.
This remarkable lack of dependency on the details allows us to understand real materials through simplified mathematical models which incorporate the same dimensions, symmetries and interaction ranges.
Furthermore, systems with disparate physics can be categorised into common universality classes.
Renormalization group\index{renormalization!group} (RG)
theory provides a satisfying, fundamental  explanation of critical behaviour and universality.
Indeed, this is one of the major achievements of statistical mechanics.

Each higher-order phase transition is characterised by a set of universal critical exponents\index{critical exponents}.
These exponents describe the strength of the phase transition in terms of power-laws.
In the 1960's, before the discovery of the renormalization group, a set of scaling relations\index{scaling relations} between these critical exponents was developed and this set is now well established and of foundational importance in the study of critical phenomena.
Because of the importance of these power laws and the associated scaling relations, circumstances where they are modified must be scrutinised and understood.
In certain situations, these power laws are modified by multiplicative logarithmic factors, which themselves are raised to certain powers.
Since 2006  a set of scaling relations between the powers of these logarithms was discovered\cite{KeJo06}, analogous to the conventional scaling relations between the leading critical exponents.  This review focuses on these universal scaling relations for logarithmic-correction exponents.\index{corrections!logarithmic}

For spin models on $d$-dimensional systems, logarithmic corrections of this type occur when the mean-field descriptions,
valid in high dimensions, turn to non-trivial power laws for smaller dimensionalities due to the importance of thermal fluctuations there.
The value of $d$ which marks the onset of the importance of these fluctuations is known as the upper critical dimension $d_c$.\index{dimension!upper critical}
Another prominent example where multiplicative logarithmic corrections are manifest is in the $q$-state Potts model\index{Potts model}
in two dimensions.
For $q \le 4$ there is a second-order phase transition there, while for $q > 4$ this transition is first order.
Logarithmic corrections arise when $q = 4$.
A more subtle example is the $d = 2$ Ising model with or without non-magnetic impurities.\index{Ising model}
In each of these cases, exponents characterise the logarithmic corrections in a manner analagous to the way in which the standard critical exponents characterise the leading power-law scaling behaviour.

The examples cited above concern well-defined Euclidean lattices.
There, the notion of space dimensionality\index{dimension} is crucial, the logarithmic corrections only arising at special values of $d$.
In recent times, the study of critical phenomena has also focused on systems defined on networks\index{networks} or random graphs.
There, instead of the system's dimensionality, a set of probabilistically-distributed coordination numbers characterised by a parameter $\lambda$ is associated with the network.
Logarithmic corrections arise at a critical value of $\lambda$.

Although logarithmic corrections to scaling are also encountered in the study of surface effects, tricritical points, the Casimir effect, and elsewhere in physics, the emphasis in this review is on bulk critical phenomena and second-order phase transitions in particular.
In the following, the leading, power-law scaling and the associated scaling relations are discussed in Sec.\ref{Second}, where the logarithmic-correction counterparts are also summarised.
The standard derivation of the scaling relations in terms of homogeneous functions and the block-spin renormalization group is recalled in Sec.\ref{Scaling}.
The logarithmic corrections are presented in Sec.\ref{Logarithmicb}, where relations between them are derived is a self-consistent manner.
Fisher renormalisation for logarithmic-correction exponents is discussed in Sec.\ref{Fren}.
In Sec.\ref{Models} the values of the correction exponents (together with the leading exponents) are given for various models exhibiting second-order phase transitions.

\section{Scaling Relations at Second-Order Phase Transitions\index{scaling relations}}
\label{Second}

Second-order phase transitions\index{phase transition!second-order} are characterised by a power-law divergence in the correlation length\index{correlation!length} $\xi$ (the length scale which describes coherent behaviour of the system). A consequence of this is the power-law behaviour of many other physical observables.
Although second-order phase transitions are also manifest in fluids, partical physics and other arenas, in this exposition we adhere  to the language of magnetism for definiteness, the translation to other fields being easily facilated.
We have in mind, then, a system of spins $s_i$ located at the sites $i$ of a $d$-dimensional lattice, whose partition function\index{partition function} is of the form
\begin{equation}
 Z = \sum_{\{s_i\}}{e^{-\beta E - h M}}.
 \label{Z}
\end{equation}
Here $E$ and $M$ represent the energy\index{energy!configurational} and magnetization\index{magnetization!configurational}, respectively, of a given configuration and the summation is over all such configurations accessible by the system. The parameters $\beta = 1/k_BT$ and $h = \beta H$ are the inverse temperature (divided by the Boltzmann constant) and the reduced external field ($H$ is the absolute strength of an applied external field).

The Helmholtz free energy\index{energy!free} is usually defined as
\begin{equation}
 F_L=-k_B T \ln{Z_L(T,H)}.
 \label{Rach}
\end{equation}
At $h=0$, a system with $N = L^d$ sites has entropy\index{entropy} given by
\begin{equation}
 Ns_L = -\frac{\partial F_L}{\partial T} = k_B \ln{Z_L} + \frac{1}{T}Ne_L,
\end{equation}
where
\begin{equation}
 Ne_L= -\frac{\partial \ln{Z_L}}{\partial \beta} = \langle{E}\rangle,
\end{equation}
is the internal energy\index{energy!internal} and $\langle{\dots}\rangle$ refer to expectation values.
Since the prefactor in Eq.(\ref{Rach}) plays no essential role in what is to come,
we drop it in the definition of the reduced free energy\index{energy!free!reduced}
\begin{equation}
 f_L = -\frac{1}{N} \ln{Z_L(T,H)}.
 \label{f}
\end{equation}
With this set-up, the reduced internal energy\index{energy!internal!reduced} and reduced entropy\index{entropy!reduced}, being the first derivative of the reduced free energy, become essentially the same.

According to a (modified) Ehrenfest\index{Ehrenfest} classification scheme, the order\index{phase transition!classification} of the transition is that of the first derivative of the free energy which displays a non-analycity in the form of a discontinuity or divergence.
For magnetically-symmetric systems this occurs at $h=0$ and at the Curie\index{Curie}\index{critical temperature} critical temperature $T_c$.
We write the reduced temperature\index{temperature!reduced} as
\begin{equation}
 t = \frac{|T-T_c|}{T_c},
 \label{t}
\end{equation}
to express the distance away from criticality in a dimensionless way.
The internal energy\index{energy!internal} $e$ and specific heat\index{specific heat} $c$ are now defined as the first and second derivatives of the  free energy with respect to temperature, respectively, while the magnetization\index{magnetization} (of the entire system) $m$ and the susceptibility\index{susceptibility} $\chi$ are defined as the first and second derivatives with respect to the external field.

\subsection{Leading Scaling Behaviour\index{scaling!leading}}
\label{Leading}

Since we are interested in behaviour near the critical point $(t,h)=(0,0)$, we write thermodynamic functions in terms of the reduced variables, {\emph{viz.}} $e_L(t,h)$, $c_L(t,h)$, $m_L(t,h)$ and $\chi_L (t,h)$, respectively. Here the subscripts indicate the size of the  system.
Then the leading, power-law, scaling behaviour\index{scaling} which describes the approach to criticality at a phase transition of second order is conventionally described as
\begin{eqnarray}
 e_\infty(t,0) \sim t^{1-\alpha} , \quad & & \quad e_\infty(0,h) \sim {h}^{\epsilon} ,
 \label{e}
\\
 c_\infty(t,0) \sim t^{-\alpha} , \quad & & \quad c_\infty(0,h) \sim {h}^{-\alpha_c} ,
 \label{C}
\\
 m_\infty(t,0) \sim t^{\beta} \quad {\mbox{for $T<T_c$,}} \quad & & \quad  m_\infty(0,h) \sim {h}^{\frac{1}{\delta}} ,
 \label{mtlead}
\\
 \chi_\infty(t,0) \sim t^{-\gamma} , \quad & & \quad  \chi_\infty(0,h) \sim {h}^{\frac{1}{\delta}-1} .
 \label{chi}
\end{eqnarray}
The numbers $\alpha$, $\beta$, $\gamma$, $\delta$, $\epsilon$ and $\alpha_c$ introduced here are called critical exponents\index{critical exponents}.

The above thermodynamic functions\index{thermodynamic functions} -- derivable from the partition function -- describe how the entire system responds to tuning the temperature and/or external field near the phase transition. To characterise local behaviour within the system we need to introduce the correlation function\index{correlation!function} and the correlation length\index{correlation!length}.
The correlation function is given by
%
\begin{equation}
 {\cal{G}}_\infty (x,t,h) \sim x^{-(d-2+\eta)}  {\tilde{\cal{G}}} \left({xt^\nu, ht^{-\Delta}}\right),
 \label{G}
\end{equation}
and the correlation length in the even (temperature) and odd (magnetic field) sectors is usually written
\begin{equation}
 \xi_\infty(t,0) \sim t^{-\nu} , \quad  \quad  \xi_\infty(0,h) \sim {h}^{-\nu_c} .
 \label{xi}
\end{equation}
Again, $\eta$, $\nu$, $\Delta$ and $\nu_c$ are critical exponents,\index{critical exponents}
with $\eta$ being called the anomalous dimension.\index{dimension!anomalous}

In addition to the above thermodynamic and correlation functions, one may  consider the zeros of the partition function\index{zeros}.
These are the complex values of the temperature or magnetic-field parameters at which the partition function vanishes.
For example,
when $h=0$, the partition function \index{partition function} in Eq.(\ref{Z}) becomes essentially a polynomial in $u = \exp {(- \beta)}$ when $L$ is finite.
Moreover, this polynomial has real coefficients.
As such, its zeros are strictly complex in the variable $u$ and appear in complex-conjugate pairs.
The zeros in the complex temperature plane at real $h$ values are denoted by $t_j(L,h)$ and are called Fisher\index{zeros!Fisher} zeros\cite{Fi65}.
The set of such zeros becomes more dense as the lattice size increases.
In the infinite-volume limit,
Fisher zeros most commonly also lie on curves in the complex temperature plane\cite{Fi65}, but may also be dispersed across 2-dimensional areas (typically this happens when anisotropy is present\cite{areas}).\index{zeros!locus}
In the thermodynamic limit they pinch the real temperature axis at the point where the phase transition occurs (namely at $t=0$).

Similarly, when $\beta$ is real and fixed,  the partition function in Eq.(\ref{Z}) becomes a polynomial in $\exp{(-h)}$.
Zeros in the complex magnetic-field plane (for real values of the reduced temperature $t$) are denoted by $h_j(L,t)$, are both $t$- and $L$-dependent and called Lee-Yang\index{zeros!Lee-Yang}zeros after their inventors\cite{LY}.
In the infinite-volume limit the Lee-Yang zeros also usually form curves in the complex plane.
In fact, in many circumstances the Lee-Yang theorem\cite{LY} ensures that these zeros are purely imaginary.

The impact of the Lee-Yang or Fisher zeros onto the real magnetic-field or temperature axis precipitates the phase transition and, in this sense, the  zeros may be considered as ``proto-critical'' points\cite{Fi78} -- they have the potential to become critical points.
Above the actual critical point $T>T_c$,  the linear locus of Lee-Yang zeros remains away from the real axis. In the thermodynamic ($L\rightarrow \infty$) limit, its lowest point is called the  Lee-Yang\index{Lee-Yang!edge} edge\cite{Fi78}  and denoted by $r_{\rm{YL}}(t)$. The edge approaches the real temperature axis as $t$ reduces to its critical value $t=0$ and that approach is also characterised by a power law.
Similarly, the lowest Fisher zero is denoted $t_1(h)$. For an infinite-sized system, this approaches the real axis as $h$ vanishes\cite{IPZ}.
The leading scaling behaviour for the edge of the distribution of Lee-Yang zeros are\cite{IPZ}
\begin{equation}
 r_{\rm{YL}}(t)  \sim t^{\Delta} , \quad  \quad  t_1(h) \sim {h}^{\frac{1}{\Delta}} .
 \label{YL}
\end{equation}

Besides the critical indices listed above, one is often interested in the so-called shift\index{shift exponent} exponent $\lambda_{\rm{shift}}$.
This characterises how the pseudocritical point in a finite-sized system is shifted away from the critical point.
The pseudocritical point\index{pseudocritical point} is the size-dependent value of the temperature $t_{\rm{pseudo}}(L)$ which marks the specific heat peak or the real part of the lowest Fisher zero.
For a system of linear extent $L$  it also scales as a power-law
\begin{equation}
 t_{\rm{pseudo}}(L)  \sim L^{\lambda_{\rm{shift}}},
 \label{shift}
\end{equation}
to leading order.
We next summarise how these critical exponents are linked through the scaling relations\index{scaling relations}.

\subsection{Scaling Relations for Leading Exponents}
\label{LeadingR}

Thus the leading power-law scaling associated with second-order phase transitions is fully described by 10 critical exponents $\alpha$, $\beta$, $\gamma$, $\delta$, $\epsilon$, $\alpha_c$, $\nu$, $\nu_c$, $\Delta$ and $\eta$  (or 11 if the exponent $\lambda_{\rm{shift}}$ characterising the finite-size scaling (FSS) of the pseudocritical point is included).\index{critical exponents}
The following 8 scaling relations (9 including one for the shift exponent) are well established for the leading critical exponents both in the (even) thermal and
(odd) magnetic sectors (see, e.g., Ref.[\refcite{Privman91}] and references therein):\index{scaling relations}
\begin{eqnarray}
 \nu d & = & 2-\alpha , \label{Jo} \\
 2\beta + \gamma & = & 2 - \alpha ,\label{Ru} \\
 \beta (\delta - 1)  & = & \gamma ,\label{Gr} \\
 \nu (2-\eta)  & = & \gamma , \label{Fi} \\
 \epsilon & = & 2 - \frac{(\delta-1)(\gamma+1)}{\delta \gamma} , \label{epsilon} \\
 \alpha_c & = & - 2 +\frac{(\gamma + 2)(\delta - 1)}{\delta \gamma} , \label{sigma} \\
  \nu_c & = & \frac{\nu}{\Delta} , \label{X} \\
 \Delta & = & \frac{\delta \gamma }{\delta - 1} = \delta \beta = \beta + \gamma \label{YLe}.
\end{eqnarray}
The relation (\ref{Jo}) was developed by Widom\cite{Wi65,Gr67} using dimensional considerations, with alternative arguments given by Kadanoff\cite{Ka66}.
Widom\cite{Wi65} also showed how  a logarithmic singularity can arise in the specific heat if $\alpha=0$, but does not have to, leaving instead a finite  discontinuity (see also Ref.[\refcite{Fi65,Gr67}]).
Later Josephson\cite{Jo67} derived a related inequality  on the basis of some plausible assumptions and  Eq.(\ref{Jo}) is sometimes called Josephson's\index{scaling relations!Josephson} relation\cite{BinneyBook}.
It can also be derived from the hyperscaling\index{scaling relations!hyperscaling} hypothesis, namely that the free energy\index{energy!free} behaves near criticality as the inverse correlation volume: $f_\infty(t,0) \sim \xi_\infty^{-d}(t)$.\index{correlation!volume}
Twice differentiating this relation recovers formula (\ref{Jo}).
For this reason, Eq.(\ref{Jo}) is also frequently called the hyperscaling relation.
It is conspicuous in the set (\ref{Jo})--(\ref{YLe}) in that it is the only scaling relation involving the dimensionality $d$.\index{dimension}

The equality (\ref{Ru}) was originally proposed by Essam and Fisher\cite{EsFi63} and a related inequality rigorously proved by Rushbrooke\cite{Ru63}.
The relation (\ref{Gr}) was advanced by Widom\cite{Wi64}, with a related inequality being proved by Griffiths\cite{Gr65}.
Equalities (\ref{Ru}) and (\ref{Gr}), sometimes called Rushbrooke's and\index{scaling relations!Rushbrooke} Griggiths'\index{scaling relations!Griffiths} laws, respectively\cite{BinneyBook}, were rederived by Abe\cite{AbeLY} and Suzuki\cite{SuzukiLY} using an alternative route involving Lee-Yang zeros.
Eq.(\ref{Fi}) was derived by\index{scaling relations!Fisher} Fisher\cite{Fi64sc}, with a related inequality proved in Ref.[\refcite{BuGu69}].
The relations (\ref{epsilon}) and (\ref{sigma}) were also derived in Refs.[\refcite{Domb,AbeLY,SuzukiLY}] and (\ref{X}) was derived in Ref.[\refcite{Ab67g}].
The reader is also referred to Ref.[\refcite{PP}].
Finally, the relationship between the gap exponent and the other exponents was established in Refs.[\refcite{Domb,AbeLY,SuzukiLY}].\index{critical exponents}

In addition to the above scaling relations, one usually finds  that\index{scaling relations}
\begin{equation}
 \lambda_{\rm{shift}} = \frac{1}{\nu}.
 \label{shiftexp}
\end{equation}
But this is not always true and in some cases, such as in the Ising model in two dimensions with special boundary conditions, it can deviate from this value\cite{JaKe02a}.
A criterion for when this may happens is given \cite{GoKe11} in Sec.\ref{shiftsec}.
It turns out that Eq.(\ref{shiftexp}) may be violated when the specific-heat amplitude ratio is 1.\index{amplitude ratio}

Because of the scaling relations, only two of the exponents listed are actually independent.\index{scaling relations}
The scaling relations (\ref{Jo})--(\ref{Fi}) are often listed as standard in textbooks, being the most frequently used.
Relations (\ref{epsilon})--(\ref{X}) involve exponents characterizing the scaling behaviour of the even thermodynamic functions $e$ and $c$ as well as of the correlation length, in field. Although less frequently encountered, their fundamental importance is similar to that of the other scaling relations.

The  formulae (\ref{e})-(\ref{shift}) characterise the {\emph{leading}} behaviour of thermodynamic and correlation functions in the vacinity of a second-order phase transition.
There are additive corrections to these scaling forms coming from both confluent\index{corrections!confluent} and analytic\index{corrections!analytic} sources.
Each scaling formula is also associated with amplitudes, so that a more complete description of the susceptibility (for example) is
\begin{equation}
 \chi(t,0) = \Gamma_\pm t^{-\gamma}
 \left({
            1 + {\mathcal{O}}(t^{\theta}) + {\mathcal{O}}(t)
 }\right),
\end{equation}
where the amplitude $\Gamma_+$ refers to the $t>0$ symmetric phase and $\Gamma_-$ corresponds to the broken-symmetry $T<T_c$ sector.
Amplitude terms such as these and additive-correction exponents are outside the remit of this review and the reader is referred to the literature\cite{BinneyBook,LeBellac}.

\subsection{Logarithmic Scaling Corrections\index{corrections!logarithmic}}
\label{LogarithmicSC}

Instead we focus on circumstances where the dominant corrections to the scaling forms (\ref{e})--(\ref{shift}) are powers of logarithms, which couple multiplicatively to the leading power laws as\index{corrections!logarithmic}
\begin{eqnarray}
e_\infty(t,0)    & \sim & t^{1-\alpha}|\ln{t}|^{\hat{\alpha}},           \label{et}\\
e_\infty(0,h)    & \sim & {h}^{\epsilon}|\ln{h}|^{\hat{\epsilon}},         \label{eh} \\
c_\infty(t,0)    & \sim & {t}^{-\alpha}|\ln{t}|^{\hat{\alpha}},            \label{ct}\\
c_\infty(0,h)    & \sim & {h}^{-\alpha_c}|\ln{h}|^{\hat{\alpha}_c},        \label{ch}\\
m_\infty(t,0)    & \sim & {t}^{\beta}|\ln{t}|^{\hat{\beta}} \quad {\mbox{for $T<T_c$}},               \label{mt}\\
m_\infty(0,h)    & \sim & {h}^{\frac{1}{\delta}}|\ln{h}|^{\hat{\delta}},   \label{mh}\\
\chi_\infty(t,0) & \sim & {t}^{-\gamma}|\ln{t}|^{\hat{\gamma}},            \label{chit}\\
\chi_\infty(0,h) & \sim & {h}^{\frac{1}{\delta}-1}|\ln{h}|^{\hat{\delta}}, \label{chih}\\
\xi_\infty(t,0)    & \sim & {t}^{-\nu}|\ln{t}|^{\hat{\nu}},                \label{xit}\\
\xi_\infty(0,h)    & \sim & {h}^{-\nu_c}|\ln{h}|^{\hat{\nu}_c},            \label{xih}\\
r_{\rm{YL}}(t) & \sim & t^{\Delta} |\ln{{t}}|^{\hat{\Delta}}
\quad {\mbox{for $t>0$}}
.
\label{edge}
\end{eqnarray}
In addition to these, the scaling of the correlation function\index{correlation!function} at $h=0$ (which is the case most often considered) may be expressed as
\begin{equation}
 {\cal{G}}_\infty (x,t,0) \sim x^{-(d-2+\eta)}(\ln{x})^{\hat{\eta}}
 D\left(
 \frac{x}{\xi_\infty(t,0)}
 \right)
 ,
 \label{corrfun}
\end{equation}
The above list of functions describe the salient features of a second-order phase transition, which is only manifest in the thermodynamic limit.
The pseudocritical-point\index{pseudocritical point} FSS is\index{shift exponent}\index{scaling!finite-size}
\begin{equation}
 t_{\rm{pseudo}}(L)  \sim L^{{\lambda}_{\rm{shift}}} (\ln{L})^{\hat{\lambda}_{\rm{shift}}},
 \label{shift2}
\end{equation}
It will turn out that the correlation length of the finite-sized system will play a crucial role and may also take logarithmic corrections, and for this reason we write\index{correlation!length}
\begin{equation}
  \xi_L(0) \sim L (\ln{L})^{\hat{q}}
  .
  \label{corrL}
\end{equation}
Note that this allows for the correlation length of the system to exceed its actual length.
For a long time this was thought not to be possible in finite-size scaling theory\cite{WatsonRuGa85}.
However we shall see that this is an essential feature of systems at their upper critical dimensionality.
Some implications of this phenomenon are discussed in Sec.\ref{Summary}.

\subsection{Scaling Relations for Logarithmic Exponents}\index{scaling relations!logarithmic}
\label{LogarithmicR}

Over the past 5 years a set of universal scaling relations for the logarithmic-correction exponents has been developed \cite{KeJo06,Pavo10,voFo11}, which connects the hatted exponents in a manner analogous to the way the standard relations (\ref{Jo})--(\ref{YLe}) relate the leading critical exponents.
While the derivation of these relations is a theme of later sections, we gather them here for convenience.  The scaling relations
for logarithmic corrections are
\begin{eqnarray}
  \hat{\alpha} & = & \left\{{\begin{array}{l}
                                             ~ 1 + d (\hat{q} -  \hat{\nu}) \quad  {\mbox{if}} \quad \alpha = 0 \quad {\mbox{and}} \quad \phi \ne \pi/4 \\
                                             ~ d (\hat{q} -  \hat{\nu})   \quad  {\mbox{otherwise,}}
                     \end{array}}\right.
  \label{SRlog1}  \\
   2 \hat{\beta} - \hat{\gamma}  & = &  d(\hat{q}-{\hat{\nu}}) ,
  \label{SRlog2}  \\
  \hat{\beta} (\delta - 1) & = &   \delta \hat{\delta} - \hat{\gamma} ,
  \label{SRlog3} \\
  \hat{\eta}  & = &  \hat{\gamma} - \hat{\nu} (2 - \eta ) ,
  \label{SRlog4} \\
  \hat{\epsilon} & = & \frac{(\gamma+1)(\hat{\beta}-\hat{\gamma} )}{\beta + \gamma} + \hat{\gamma},
  \label{SRlog5} \\
  \hat{\alpha}_c & = &  \frac{(\gamma+2)(\hat{\beta}-\hat{\gamma} )}{\beta + \gamma}+ \hat{\gamma},
  \label{SRlog6} \\
  \hat{\delta}  & = & d \hat{q} - d \hat{\nu}_c ,
  \label{SRlogX}\\
  \hat{\Delta}  & = & \hat{\beta} - \hat{\gamma},
  \label{SRlog7}\\
  \hat{\lambda}_{\rm{shift}}  & = & \frac{\hat{\nu} - \hat{q}}{\nu}.
  \label{SRlog8}
\end{eqnarray}
In the first of these, $\phi$ refers to the angle at which the complex-temperature zeros impact onto the real axis.
If $\alpha = 0$, and if this impact angle\index{impact angle} is any value other than $\pi/4$, an extra logarithm arises in the specific heat.
This is expected to happen in $d=2$ dimensions, but not in $d=4$, where $\phi = \pi/4$ \cite{KeJo06}.
The static scaling relations (\ref{SRlog2}), (\ref{SRlog3}), (\ref{SRlog5}), (\ref{SRlog6}), (\ref{SRlog7}), and (\ref{SRlog8}) can be deduced from the Widom scaling hypothesis that the Helmholtz free energy is a homogeneous function\cite{Wi65} but the others require more careful deliberations, as we shall see.
\index{scaling!static}\index{scaling!Widom}

\section{Standard Derivation of Leading Scaling Relations}
\label{Scaling}

In this section we show how the leading-power-law scaling relations (\ref{Jo})--(\ref{Fi}) are derived using the Widom scaling hypothesis\index{scaling!Widom} and Kadanoff's block-spin approach.
The remaining leading-power-law  relations (\ref{epsilon})--({\ref{X})
are derived in a similar manner.
The presentation here is necessarily elementary and the reader is refered to the standard literature (e.g., Ref.[\refcite{LeBellac}]) for more in-depth treatments.
The derivation of relation (\ref{YLe}) is reserved for the subsequent section where the emphasis is on partition function zeros.

\subsection{Static Scaling}
\label{Static}

The Widom (or static) scaling hypothesis\index{scaling!static} is that the free energy $f$ (or at least its singular part, which is responsible for divergences in the thermodynamic functions at the critical point) is a generalized homogeneous function (see Appendix~\ref{Ahomogeneous}).
It is convenient to  express Widom homogeneity as
\begin{equation}
 f(t,h) =  b^{-d} f(b^{y_t} t, b^{y_h} h)
 ,
 \label{Widomhyp}
\end{equation}
where $b$ is a dimensionless rescaling parameter.
In the renormalization group (RG) context, $t$ and $h$ are called {\emph{linear scaling fields}} and $y_t$ and $y_h$ are {\emph{RG eigenvalues.}}
Differentiating Eq.(\ref{Widomhyp}) with respect to $h$ gives the magnetization and susceptibility as
\begin{equation}
 m(t,h) = b^{y_h-d} m(b^{y_t} t, b^{y_h} h)
 \quad {\mbox{and}}
 \quad  \chi(t,h) = b^{2y_h-d} m(b^{y_t} t, b^{y_h} h),
 \label{mth}
\end{equation}
while  appropriate differentiation with respect to $t$ gives
the internal energy (entropy) and specific heat,
\begin{equation}
 e(t,h) =  b^{y_t-d} f(b^{y_t} t, b^{y_h} h)
  \quad {\mbox{and}} \quad
 c(t,h) =  b^{2y_t-d} f(b^{y_t} t, b^{y_h} h)
 .
 \label{cth}
\end{equation}
At $h=0$, the choice $b = t^{-1/y_t}$ recovers the first expressions in each of Eqs.(\ref{e})--(\ref{chi}) provided
\begin{equation}
 \alpha = \frac{2y_t-d}{y_t},
 \quad
 \beta = \frac{d-y_h}{y_t}
 \quad {\mbox{and}} \quad
 \gamma = \frac{2y_h-d}{y_t}
 .
\label{alpha1}
\end{equation}
On the other hand, at $t=0$, the choice $b = h^{-1/y_h}$ recovers the remaining parts of Eqs.(\ref{e})--(\ref{chi}) if
\begin{equation}
 \delta = \frac{y_h}{d-y_h}
, \quad
 \epsilon = \frac{d-y_t}{y_h}
 \quad {\mbox{and}} \quad
 \alpha_c = \frac{2y_t-d}{y_h}
 .
\label{delta1}
\end{equation}
The  scaling hypothesis therefore allows one to express
the six static critical exponents $\alpha$, $\beta$, $\gamma$, $\delta$, $\epsilon$ and $\alpha_c$
in terms of $y_t/d$ and $y_h/d$. In particular,
\begin{equation}
 \frac{y_t}{d} = \frac{1}{\beta(\delta+1)}
 , \quad \quad
 \frac{y_h}{d} = \frac{\delta}{\delta+1}
 .
\label{ytyh}
\end{equation}
These can now be eliminated from the remaining equations
(\ref{alpha1}) and (\ref{delta1}) to give scaling relations (\ref{Ru}), (\ref{Gr}), (\ref{epsilon}) and (\ref{sigma}).

The hyperscaling scaling relation (\ref{Jo}) and Fisher's relation (\ref{Fi}) are of rather a different status than the others in that they involve the exponents $\nu$ and $\eta$, which are associated with
local rather than global properties of the system.
To derive these, we need more than the Widom homogeneous scaling hypothesis.\index{scaling relations!hyperscaling}\index{scaling relations!Fisher}

\subsection{Renormalization Group}
\label{Renormalization}

To illustrate the renormalization group\index{renormalization!group}, we have in mind $N=L^d$  Ising spins, for example, on a $d$-dimensional lattice with spacing $a$ (Fig.\ref{figRK1}). The Hamiltonian for the system is
\begin{equation}
 {\mathcal{H}} = -J \sum_{\langle{i,j}\rangle}{s_i s_j} - H \sum_i{s_i}
,
\label{Ising}
\end{equation}
where $s_i \in \{\pm 1\}$ is an Ising spin at the $i$th lattice site and where the interaction is between nearest neighbours.
In anticipation of the renormalization group (RG), we generalise this by introducing other locally interacting terms
such as sums over plaquettes,
next-nearest-neighbour interactions, etc., and
we consider a reduced Hamiltonian
\begin{equation}
 \bar{{\mathcal{H}}} = \beta {\mathcal{H}} = -K_t \sum_{\langle{i,j}\rangle}{s_i s_j} -
 K_h \sum_i{s_i}
 - K_3 \sum_{\langle{i,j,k,l}\rangle}{s_i s_j s_k s_l}
 - K_4 \sum_{\ll{i,j}\gg}{s_i s_j}
  - \dots
,
 \end{equation}
 where $K_t=\beta J$, $K_h = h$, and the remaining $K_n$
 are similar coupling strengths.
 Here, all of the additional terms have to be symetrical under $s_i \rightarrow s_j$ and
 $\ll \dots \gg $ indicates interactions between next-nearest neighbours.

\begin{figure}
\centerline{\psfig{file=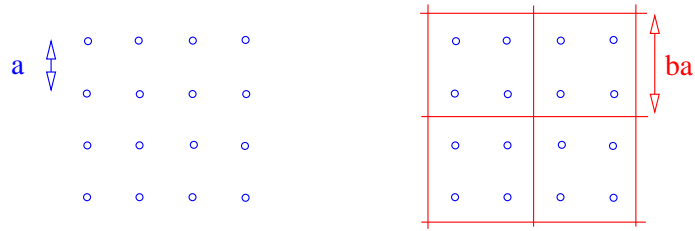,width=9.0cm}}
\caption{In the block-spin process, the original lattice with spacing $a$ and $N$ spins becomes one of spacing $ba$ with $N/b^d$ block spins.}
\label{figRK1}
\end{figure}

As illustrated in Fig.\ref{figRK1}, we place the spins into blocks\index{block spins} of length $ba$ and rescale the $N^\prime = N/b^d$ block spins so they each have magnitude 1,
\begin{equation}
 s^\prime_I = {\mathcal{F}}\left({\{s_i\}_{i \in I}}\right) = \frac{c(b)}{b^d} \sum_{i\in I}{s_i}
,
\label{SI}
\end{equation}
where $s^\prime_I$ labels an Ising block spin at site $I$ of the new lattice.
Here $c(b)$ is a spin dilatation factor. (If all spins in a block were fully aligned then $c(b)$ would be $1$.)
We require that the partition function of the blocked system is the same as the original one:
\begin{equation}
 Z_{N^\prime}(\bar{{\mathcal{H}}^\prime}) = Z_{N}(\bar{{\mathcal{H}}})
  .
 \label{ZZ}
\end{equation}
This requirement induces a transformation from the point
 ${\vec{\mu}} = (K_1,K_2,\dots)$ in the parameter space to another point ${\vec{\mu}}^\prime = (K_1^\prime,K_2^\prime,\dots)$ and we write ${\vec{\mu}}^\prime = R_b {\vec{\mu}}$.
The RG approach presumes the existence of a fixed point $\vec{\mu}^*$ which is invariant under $R_b$. If ${\vec{\mu}} = \vec{\mu}^* + \delta \vec{\mu}$ and
${\vec{\mu}}^\prime = \vec{\mu}^* + \delta \vec{\mu}^\prime$ are nearby this fixed point, we linearise the RG by a Taylor expansion,
\begin{equation}
 \vec{\mu}^* + \delta \vec{\mu}^\prime
 =
 R_b(\vec{\mu}^* + \delta \vec{\mu})
 =
 \vec{\mu}^* + R_b^\prime(\vec{\mu}^*) \delta \vec{\mu} + \dots,
\end{equation}
so that
\begin{equation}
 \delta \vec{\mu}^\prime = R_b^\prime(\vec{\mu}^*) \delta \vec{\mu}
 .
\label{Yeo8.17}
\end{equation}
Let the eigenvectors and eigenvalues of $R_b^\prime(\vec{\mu}^*)$
be given by
\begin{equation}
 R_b^\prime(\vec{\mu}^*) \vec{v}_i = \lambda_i(b) \vec{v}_i
 .
\end{equation}
Demanding that two successive applications of the RG transformation using scale factors $b_1$ and $b_2$ are equivalent to a single transformation with scale factor $b_1b_2$, we have that $\lambda_i(b_1b_2) =
\lambda_i(b_1)\lambda_i(b_2)$, so that  $\lambda_i$ is a homogeneous function and (see Appendix~\ref{Ahomogeneous})
\begin{equation}
 \lambda_i(b) = b^{y_i}
 .
 \label{Yeo8.19}
\end{equation}
Vectors $\vec{\mu}$ and $\vec{\mu}^*$ may be expanded in terms of the eigenvectors $\vec{v}_i$:
\begin{equation}
 \delta \vec{\mu} =  \sum_i{t_i \vec{v}_i}
 ,
 \quad \quad
 \delta \vec{\mu}^\prime =  \sum_i{t_i^\prime \vec{v}_i}
 .
\label{Yeo8.20}
\end{equation}
The $t_i$ here are called linear scaling fields.
Eq.(\ref{Yeo8.19}) then gives the linear version of Eq.(\ref{Yeo8.17}) to be
\begin{equation}
 \sum_i{t_i^\prime \vec{v}_i}  = \sum_i{t_i b^{y_i} \vec{v}_i},
\end{equation}
and the linear scaling fields transform under the RG as
\begin{equation}
 t_i^\prime = b^{y_i}t_i
 .
 \label{Yeo8.22}
\end{equation}

There are three cases to distinguish:
\begin{itemize}
\item
If $y_i >0$, the associated linear scaling field\index{scaling!field} $t_i$ is called relevant\index{scaling!field!relevant} -- it increases away from the fixed point under successive applications of the RG.
\item
If $y_i <0$, the scaling field decreases towards the fixed point under the RG and is called irrelevant\index{scaling!field!irrelevant}.
 \item
If $y_i=0$, the field $t_i$ is termed marginal\index{scaling!field!marginal}. In this critical exponents may be continuously dependent on the parameters of the Hamiltonian or  logarithmic corrections may arise.
\end{itemize}

Since the partition functions (\ref{ZZ}) of the original and renormalized systems are the same, the free energies near the fixed point are related by
\begin{equation}
 N^\prime f_{N^\prime}(\bar{{\mathcal{H}}^\prime}) = N f_{N}(\bar{{\mathcal{H}}})
 ,
\end{equation}
or, since $N^\prime = N/b^d$,
\begin{equation}
 f(t_1,t_2,t_3,\dots) = b^{-d} f(t_1^\prime,t_2^\prime,t_3^\prime,\dots)
 =
 b^{-d} f(b^{y_1}t_1,b^{y_2}t_2,b^{y_3}t_3,\dots)
 .
\end{equation}
This is precisely the Widom scaling form (\ref{Widomhyp}) leading to the
static scaling relations.\index{scaling!static}

To investigate local properties, consider the correlation function of the block spins,\index{block spins}
\begin{equation}
 G(r_{IJ}) = \langle{ s^\prime_I s^\prime_J }\rangle
 =
 \frac{1}{Z}
 \sum_{\{{s^\prime}\}} s^\prime_I s^\prime_J e^{-\bar{\mathcal{H}^\prime}(s^\prime)}
 ,
 \label{SISJ}
\end{equation}
where $r_{IJ}$ is the distance between block $I$ and block $J$ in units of the block-spin distance. The sum is over the block-spin configurations.
Now, for a given $s^\prime_I$ value at a block-spin-site $I$, there is a multitude of sub-configurations for the set $\{ s_i \}$ with $i \in I$. The probability that the $I$th blocked spin takes a particular value (say $s^\prime_I=1$) is then
${\displaystyle{\sum_{\{s_i\}_{i \in I}}{ \delta(s^\prime_I - 1) e^{-\mathcal{H}(s)}}}}/Z$.
Similarly, the probability that the blocked system is in any particular configuration given by Eq.(\ref{SI}) is
$e^{-\mathcal{H}(s)}/Z$ where
\begin{equation}
e^{-{\mathcal{H}}^\prime{(s^\prime)}}
=
\sum_{\{s\}}{ \prod_I{\delta(s^\prime_I - {\mathcal{F}}\left({\{s_i\}_{i \in I}}\right)) e^{-\mathcal{H}(s)}}} .
\end{equation}
Putting this expression into Eq.(\ref{SISJ}), we obtain
\begin{eqnarray}
 G_{IJ} = \langle{ s^\prime_I s^\prime_J }\rangle
 & = &
 \frac{1}{Z}
 \sum_{\{{s}\}} s^\prime_I s^\prime_J { \prod_K{\delta(s^\prime_K - {\mathcal{F}}\left({\{s_k\}_{k \in K}}\right) e^{-\mathcal{H}(s)}}}
 \\
  & = &
  \frac{1}{Z} \frac{c^2(b)}{b^{2d}}
  \sum_{\{{s}\}}
  \left({\sum_{i \in I}{s_i}}\right)
  \left({\sum_{j \in J}{s_j}}\right)
   e^{-\mathcal{H}(s)}
    \\
  & = &
  \frac{c^2(b)}{b^{2d}}
  {\sum_{i \in I}}{\sum_{j \in J}{\left\langle{s_is_j}\right\rangle}}
    \\
  & \approx &
   c^2(b) {\left\langle{s_is_j}\right\rangle}
    =
    c^2(b) G_{ij}
  ,
\end{eqnarray}
provided the two blocks $I$ and $J$ are sufficiently far apart that
${\left\langle{s_{i^\prime}s_{j^\prime}}\right\rangle} \approx {\left\langle{s_is_j}\right\rangle} $ for
$i,i^\prime \in I$ and $j, j^\prime \in J$.
If
\begin{equation}
 G_{ij} \sim e^{-r_{ij}/\xi}
 ,
\end{equation}
then since $r_{IJ}=r_{ij}/b$, we have the  property that
\begin{equation}
 \xi^\prime = \frac{\xi}{b}
 ,
 \label{nice}
\end{equation}
where $\xi^\prime$ is the block-spin correlation length,
and that
\begin{equation}
 G\left({\frac{r}{b}, \mu^\prime}\right)
  \approx
 c^2(b)
 G\left({r, \mu}\right)
.
\end{equation}
Demanding as usual that $c(b_1b_2) = c(b_1)c(b_2)$ (homogeneity) means
a power law, which we write as
\begin{equation}
 c(b) = b^{d_\phi}
.
\end{equation}
Here $d_\phi$ is called the {\emph{anomalous dimension of the field.}}\index{dimension!anomalous}
The choice $b=r$ gives that $G(r) \sim c^{-2}(r) \sim r^{-2d_\phi}$.
Comparing this to Eq.(\ref{G}), we obtain
\begin{equation}
 d_\phi = \frac{1}{2}(d-2+\eta)
.
\label{dphi}
\end{equation}

Eq.(\ref{Yeo8.22}) with $t_i=t$ and $y_i=y_t$ gives $t^\prime = b^{y_t}t$.
Alternatively, $t_i=h$ and $y_i=y_h$ gives $h^\prime = b^{y_h}h$.
From Eq.(\ref{nice}), we may write these as
$ t \xi^{y_t} = t^\prime {\xi^\prime}^{y_t} $
or
$ h \xi^{y_h} = h^\prime {\xi^\prime}^{y_h} $.
Therefore we identify $\xi \sim t^{-1/ y_t}$ or $\xi \sim h^{-1/ y_h}$,
depending on whether $h=0$ or $t=0$. From Eq.(\ref{xi}),
these give
\begin{equation}
 \nu = \frac{1}{y_t} \quad  {\mbox{and}} \quad \nu_c = \frac{1}{y_h}.
\label{LB321}
\end{equation}
Now, $y_t$ is related to the critical exponent $\alpha$ through Eq.(\ref{alpha1}). Combining this with Eq.(\ref{LB321}) we obtain the hyperscaling\index{scaling relations!hyperscaling} relation (\ref{Jo}). Eq.(\ref{ytyh}) then leads to Eq.(\ref{X}) for the exponent $\nu_c$.

In summary, the scaling relations (\ref{Ru}), (\ref{Gr}) (\ref{epsilon}) and (\ref{sigma}) may be deduced from the Widom scaling hypothesis \cite{Wi65}, and Eqs.(\ref{Jo}) and (\ref{X}) can be derived from the Kadanoff block-spin construction \cite{Ka66} and ultimately from Wilson's RG \cite{Wi71}.
Next we come to Fisher's scaling relation (\ref{Fi}). There are at least two ways to do this in the literature, one more common than the other. We describe these in turn.

\subsection{Fisher's  Scaling Relation}\index{scaling relations!Fisher's relation}
\label{Fishers}

With $h=0$, fixing the remaining argument of the correlation function (\ref{G}), we may write
\begin{equation}
 {\cal{G}}_\infty (x,t,0)  \sim
 \xi_\infty(t,0)^{-(d-2+\eta)}
 D\left( \frac{x}{\xi_\infty(t)} \right)
.
\label{AH1}
\end{equation}
Following the original approach used by Fisher \cite{Fi64sc}, and writing the susceptibility as
\begin{equation}
 \chi_\infty(t,0) = \int_0^{\xi_\infty(t,0)}{d^dx {\cal{G}}_\infty (x,t)}
,
\label{Fishertrick}
\end{equation}
one obtains
\begin{equation}
 \chi_\infty(t,0) \sim \xi_\infty(t,0)^{2-\eta}
.
\label{wind}
\end{equation}
The leading power-laws in Eqs.(\ref{chi}) and (\ref{xi}) then deliver the scaling relation (\ref{Fi}).

An alternative argument  presented in the literature \cite{SaSo97}
 (see also Ref.[\refcite{Ab67g,Su68}])
attempts to relate the correlation function to the magnetization.\index{correlation function}\index{magnetization}
If the spins decorrelate  in the large-distance limit, one may expect that
\begin{equation}
 {\cal{G}}_\infty (x,t) = \langle {\vec{s}}(0){\vec{s}}(x)\rangle
\rightarrow \langle {\vec{s}}(0)\rangle \langle {\vec{s}}(x)\rangle
= m_\infty^2(t,0)
\label{green}
\end{equation}
there.
Using Eqs.(\ref{G}) and (\ref{xi}) for the left hand side and
matching with Eq.(\ref{mtlead}) on the right, one again obtains
the standard scaling relation  (\ref{Fi}) from the leading exponents.
Although this approach delivers the correct result in this instance, we will later see that the technique
delivers a different result at the logarithmic level in general.

\subsection{The Shift Exponent}\index{shift exponent}
\label{shiftsec}

The scaling relation for the shift exponent was given in Eq.(\ref{shiftexp}) as the inverse of the correlation-length exponent $\nu$.
This is immediately derived from Eq.(\ref{Widomhyp}) which, with $b=L$ and in vanishing field,  implies
\begin{equation}
 c_L(t,0) = L^{-d}c(L^{\frac{1}{\nu}} t,0),
 \label{ZQ}
\end{equation}
having used Eq.(\ref{LB321}). Setting the temperature derivative to zero (to maximise the specific heat) gives that the specific-heat peak scales as in Eq.(\ref{shift}) with $\lambda_{\rm{shift}}$ given by Eq.(\ref{shiftexp}).
However, it was also pointed out in Sec.\ref{LeadingR} that this relation does not always hold.
One may ask whether there is a criterion for which to decide on the validity of Eq.(\ref{shiftexp}).
Such a criterion was recently derived\cite{GoKe11}.

Although there is no FSS\index{scaling!finite-size} theory for the impact angle\index{impact angle} of Fisher zeros, for sufficiently large lattice size $L$ one may expect that $\phi$ is approximated by the angle subtended by the lowest-lying zero on the real axis at the critical point,
\begin{equation}
  \tan \phi \approx \frac{{\rm{Im}} [t_1(L)]}{{\rm{Re}}[ t_1(L)]}
  \propto \frac{L^{-1/\nu}}{L^{-\lambda_{\rm{shift}}}} \sim C L^{\lambda_{\rm{shift}} - 1/\nu} + \dots.
\end{equation}
Here we have used the FSS result that\cite{IPZ} ${\rm{Im}} [t_1(L)]\sim L^{-1/\nu}$.
The angle $\phi$ is given by $L \rightarrow \infty$. If, as in most cases, $\lambda_{\rm{shift}} = 1/\nu$, then $\tan{\phi}$ is a finite value.
If $\lambda_{\rm{shift}} < 1/\nu$ then $\tan \phi = 0$. But it is impossible for $\lambda_{\rm{shift}}$ to be less than $1/\nu$ -- otherwise $\nu$ \emph{becomes} $1/\lambda_{\rm{shift}}$.
The only alternative to $\lambda_{\rm{shift}} = 1/\nu$, then, is $\lambda_{\rm{shift}} > 1/\nu$.
In this case  $\tan \phi$ diverges as $L \rightarrow \infty$, so that  $\phi = \pi/2$ and the zeros impact onto the real axis vertically.
Vertical impact implies a symmetry between the low- and high-temperature phases, which in turn implies that the specific heat amplitudes on either side of the critical point must coincide.
In other words, only in circumstances where the specific-heat amplitude ratio\index{amplitude ratio} (a universal\index{universality} quantity) is unity  is it allowed to violate Eq.(\ref{shiftexp}).

In the Ising  model in $d=2$ dimensions, vertical impact of the Fisher zeros and the coincidence of the specific-heat amplitudes is guaranteed by self-duality.
In this case, Ferdinand and Fisher found $\lambda_{\rm{shift}} = 1/\nu = 1$ for the square lattice Ising model with periodic boundaries \cite{ferdinand1969}, but other two-dimensional lattices with different topologies have different values of $\lambda_{\rm{shift}}$\cite{JaKe02a}.

\section{Logarithmic Corrections}
\label{Logarithmicb}

Logarithmic corrections\index{corrections!logarithmic} are characteristic of marginal scenarios\index{scaling!field!marginal}  (see, e.g., Ref.[\refcite{We76}] and references therein).
The hyperscaling hypothesis\index{scaling relations!hyperscaling} $f_\infty(t,0) \sim \xi_\infty^{-d}(t,0)$ fails at and above the upper critical\index{dimension!critical} dimension $d_c$. While the leading scaling relation (\ref{Jo}) holds at $d_c$ itself, it fails for $d>d_c$, where mean-field behaviour  prevails. This mean-field behaviour holds independent of $d$ (provided $d>d_c$), so the critical exponents and scaling relations should also be $d$-independent there. At $d_c$ itself, multiplicative logarithmic corrections to scaling appear.

The two-dimensional $q$--state Potts model\index{Potts model} has a first-order phase transition for $q > 4$ and a second-order one when $q \le 4$.
The borderline $q=4$ case manifests  multiplicative logarithmic corrections to scaling.
The $q=2$ version of the Potts model is the Ising model.\index{Ising model} In two dimensions this has a logarithmic divergence in the specific heat. Although no other thermodynamic quantity manifests such a logarithm in this model, this scenario also has to be accounted for in any theory of logarithmic corrections.

Staying in two dimensions, the Ising model with uncorrelated, quenched, random-site or bond disorder is another example where logarithmic corrections appear at a demarcation point.
The Harris\index{Harris} criterion\cite{Ha74} tells us that when the critical exponent $\alpha$ of a pure system is positive, random quenched disorder\index{disorder} is relevant \cite{Ha74}. This means that critical exponents may change as disorder is added to the system.  If $\alpha$ is negative in the pure system, the critical behaviour is not changed by such disorder. In the marginal $\alpha = 0$, no Harris  prediction is possible, and there logarithmic corrections to the pure model  may ensue.

Because of the ubiquity of these logarithms in critical phenomena, it
is reasonable to seek scaling relations for their exponents in analogy to Eqs.(\ref{Jo})--(\ref{YLe}) and (\ref{shiftexp}) above.
These are the  scaling relations of Sec.\ref{LogarithmicSC}, which have only recently been developed \cite{KeJo06,Pavo10,voFo11}.
Here, these theoretical developments are brought together and summarised, and their consequences are confronted with the literature.
In many cases the values of logarithmic corrections derived in the literature using a multitude of disparate techniques are upheld. A few cases which conflict with the literature are highlighted as requiring further investigations. Finally, holes in the literature are filled, pointing the way for further research endeavours into the future.

The scaling theory presented in this section is entirely based on self-consistencies -- independent of RG. The theory does not {\emph{predict}} the existence of logarithmic corrections; rather, when they are known by other methods to be present, the theory restricts their form through scaling relations.
For ab inito model-specific theories, the RG  is more appropriate
and the reader is referred to the literature\cite{Wi71,Wegnerlog1,Wegnerlog2,approp,history}.
Wegner's analysis\cite{Wegnerlog1}, in particular, uncovered the role of marginal variables and nonlinear scaling fields.
These were further developed by Huse and Fisher\cite{approp}.
Excellent reviews on the origins of logarithmic corrections in the paradigmatic four-state Potts case are contained in Ref.[\refcite{ShchurBerche}].

\subsection{Static Correction Exponents}
\label{StaticC}

We denote the $j$th Lee-Yang zero\index{zeros!Lee-Yang} for a system of size $L$ by
\begin{equation}
h_j(t,L) = r_j(t,L) \exp{(i \phi(r_j(t,L)))} .
\end{equation}
Here $h_j(t,L) $ is complex and $r_j(t,L)$ is real. The latter parameterises the position of the zero along the locus of all zeros which is given by the function $\phi(r)$. If the Lee-Yang theorem\index{theorem!Lee-Yang} holds \cite{LY}, the angle $\phi(r)$ is $\pi/2$ and the zeros are on the imaginary field axis.
In fact the validity of the Lee-Yang  theorem is not required or assumed in what follows (and it does not hold for the Potts model, for example).
Expressed in terms of these zeros, the finite-size partition function is
\begin{equation}
 Z_L(t,h) \propto \prod_j{(h-h_j(t,L))} ,
\label{Zitozeros}
\end{equation}
where the product is over all zeros and constant of proportionality (which is not displayed) contributes only to the regular part of the free energy.
The logarithm in the free energy converts the product into a sum and
\begin{equation}
 f_L(t,h) =  \frac{1}{N}\sum_j{\ln{(h-h_j(t,L))}} ,
\label{fitozeros}
\end{equation}
up to an additive constant (not displayed). Defining
\begin{equation}
 g_L(r,t) = \frac{1}{N} \sum_j{\delta{(r - r_j(t,L))}},
\label{densityofzerosN}
\end{equation}
we may write
\begin{equation}
 f_L (t,h) = \int{\ln{(h-h(r,t))}g_L(r,t) dr }
 ,
\label{g0}
\end{equation}
where the integral is over the locus of zeros\index{zeros!locus}. Since these occur in complex conjugate pairs, and since the density of zeros vanishes up to $r = r_{\rm{YL}}(t)$, we express   the free energy\index{energy!free} in the
thermodynamic limit as
\begin{equation}
 f_\infty (t,h) = 2 {\rm{Re}} \int_{r_{\rm{YL}}(t)}^R{\ln{(h-h(r,t))}g_\infty(r,t) dr }
 .
\label{g1}
\end{equation}
We have assumed that critical behaviour is dominated by the Lee-Yang zeros closest to the critical point and that the locus of these zeros can be approximated by $\phi(r,t)  = \phi$, which is a constant.
We have also inserted an integral cutoff $R$.

The  susceptibility\index{susceptibility} is the second  derivative of the free energy with respect to the external field. It is convenient to substitute $r = xr_{\rm{YL}}(t)$, so that at $h=0$  it is
\begin{equation}
 \chi_\infty(t,0) = -\frac{
                       2\cos{(2\phi)}
                      }{
                       r_{\rm{YL}}(t)
                      }
\int_1^{\frac{R}{
                 r_{\rm{YL}}(t)
                 }}{
                            \frac{g_\infty(xr_{\rm{YL}},t)}{x^2}
                         dx }
 .
\label{g4}
\end{equation}
Expanding Eq.(\ref{g4}) about $r_{\rm{YL}}(t)/R=0$, which is reasonable near criticality, one finds
\begin{equation}
 g_\infty(r,t) =
 \chi_\infty(t,0) r_{\rm{YL}}(t)
 \Phi{\left(\frac{r}{r_{\rm{YL}}(t)}\right)}
,
\label{g5}
\end{equation}
up to additive corrections in $r_{\rm{YL}}(t)/R$ and
where $\Phi$ is an undetermined function.
The ratio $r_{\rm{YL}}(t)/R$ is assumed small enough near criticality to drop  additive corrections.
Analogous deliberations for the magnetization\index{magnetization} in field give
\begin{equation}
 m_\infty(t,h) =
 \chi_\infty(t,0)  r_{\rm{YL}}(t)
\Psi_\phi{\left(\frac{h}{r_{\rm{YL}}(t)}\right)}
  ,
\label{g7}
\end{equation}
where
\begin{equation}
 \Psi_\phi{\left(\frac{h}{r_{\rm{YL}}(t)}\right)}
=
2 {\rm{Re}}
\int_1^\infty{\frac{\Phi(x) }{h/r_{\rm{YL}}(t)-xe^{i\phi}} dx }
 .
\label{g6}
\end{equation}
Now, allowing $h \rightarrow 0$ in Eq.(\ref{g7}), and comparing to 
the assumed scaling form of Eq.(\ref{mt}), one recovers the leading scaling relation (\ref{YLe}). One also arrives at the first scaling relation for logarithmic corrections, namely
\begin{equation}
 \hat{\Delta}  =  \hat{\beta} - \hat{\gamma}
.
\label{NewLY}
\end{equation}
Furthermore, fixing the argument of the function $\Psi_\phi$ in Eq.(\ref{g7}) leads to  $t \sim h^{1/\Delta}|\ln{{h}}|^{-\hat{\Delta}/\Delta}$ using Eq.(\ref{edge}),
so that Eq.(\ref{g7}) may be expressed as
\begin{equation}
 m_\infty(t,h)
 \sim h^{1-\frac{\gamma}{\Delta}}
|\ln{h}|^{\hat{\gamma}+\frac{\gamma \hat{\Delta}}{\Delta}}
\Psi_\phi{\left(\frac{h}{r_{\rm{YL}}(t)}\right)}
  .
\label{g77}
\end{equation}
Next taking the limit $t \rightarrow 0$ and comparing with the assumed form
Eq.(\ref{mh}), recovers the leading edge behaviour (\ref{YLe}).
It also delivers the correction relation $\hat{\Delta} = \delta (\hat{\delta}-\hat{\gamma})/(\delta - 1)$.
The former recovers the leading scaling relation (\ref{Gr}), while the latter, together with Eq.(\ref{NewLY}), gives its logarithmic counterpart,
\begin{equation}
 \hat{\beta} (\delta - 1) =  \delta \hat{\delta} - \hat{\gamma}
.
\label{NewW1}
\end{equation}
Note that the assumption of logarithmic corrections to the zeros in Eq.(\ref{edge}) necessarily leads to logarithms
in the other observables.
If one tries to omit them in other quantities (or vice versa), contradictions ensue.
E.g., attempting to force $\hat{\gamma}=0$ still necessitates a logarithmic correction to the magnetization
through a non-zero value of $\hat{\Delta}$ in Eq.(\ref{g77}).
 Alternatively, attempting to force $\hat{\beta}=0$ necessitates a non-vanishing $\hat{\gamma}$.
By the same token, omission of logarithmic correction factors is the same as setting the relevant hatted exponents to zero
and leads to contradictions.
Therefore allowing for logarithmic corrections\index{corrections!logarithmic} from the start reflects the philosophy of this exposition,
which is based upon self-consistencies,
rather than the Wegner\index{corrections!Wegner} approach \cite{Wegnerlog1,Wegnerlog2} which predicts their existence.

We next introduce the cumulative distribution function of zeros:
\begin{eqnarray}
 G_\infty (r,t)
 & = &
 \int_{r_{\rm{YL}}(t)}^{r}{g_\infty(s,t)ds}
\label{Gdensity}
 \\
 & = &
 \chi_\infty(t,0) r^2_{\rm{YL}}(t)
 I{\left(\frac{r}{r_{\rm{YL}}(t)}\right)}
,
\label{GG}
\end{eqnarray}
where $I(y) = \int_1^y{\Phi(z)dz}$.
Applying integration by parts to the free energy  (\ref{g1}) then gives  its singular part to be
\begin{equation}
 f_\infty(t,h) = -2 {\rm{Re}} \int_{r_{\rm{YL}}(t)}^R{\frac{G_\infty(r,t) \exp{(i\phi)} dr}{h - r \exp{(i\phi)}}}
.
\label{rain}
\end{equation}
Again substituting $r=xr_{\rm{YL}}(t)$,
\begin{equation}
  f_\infty(t,h) = \chi_\infty(t,0) r_{\rm{YL}}^2(t) {\cal{F}}_\phi
 \left({ \frac{h}{r_{\rm{YL}}(t)} }\right)
,
\end{equation}
where
\begin{equation}
 {\cal{F}}_\phi (y) = -2 {\rm{Re}}\int_1^\infty{\frac{I(x)dx}{y \exp{(-i\phi)}-x}}
,
\end{equation}
and we have taken the limit $R/{\rm{YL}}(t) \rightarrow \infty$.
The internal energy\index{energy!internal} and specific heat\index{specific heat} are given by
the first and second derivatives of the free energy with respect to $t$, respectively. These are
\begin{eqnarray}
  e_\infty(t,h) & = & \chi_\infty(t,0) r_{\rm{YL}}^2(t) t^{-1} {\cal{F}}_\phi
 \left({ \frac{h}{r_{\rm{YL}}(t)} }\right)
,
\label{eYurko}
\\
  c_\infty(t,h) & = & \chi_\infty(t,0) r_{\rm{YL}}^2(t) t^{-2} {\cal{F}}_\phi
 \left({ \frac{h}{r_{\rm{YL}}(t)} }\right)
.
\label{cYurko}
\end{eqnarray}
Inverting Eq.(\ref{edge}), we may write
\begin{equation}
t \sim r_{\rm{YL}}^{\frac{1}{\Delta}} |\ln{r_{\rm{YL}}}|^{-\frac{\hat{\Delta}}{\Delta}}
,
\end{equation}
which with Eq.(\ref{chit}) gives
 \begin{equation}
\chi_\infty(t,0) \sim r_{\rm{YL}}^{-\frac{\gamma}{\Delta}} |\ln{r_{\rm{YL}}}|^{\frac{\gamma \hat{\Delta}}{\Delta} + \hat{\gamma}}
.
\end{equation}
Together, these give
\begin{eqnarray}
  e_\infty(t,h) & = & r_{\rm{YL}}^{2-\frac{\gamma}{\Delta}-\frac{1}{\Delta}} |\ln{r_{\rm{YL}}}|^{\frac{(\gamma+1)\hat{\Delta}}{\Delta}+\hat{\gamma}}  {\cal{F}}_\phi   \left({ \frac{h}{r_{\rm{YL}}} }\right)
,
\label{e1}
\\
 c_\infty(t,h) & = & r_{\rm{YL}}^{2-\frac{\gamma}{\Delta}-\frac{2}{\Delta}} |\ln{r_{\rm{YL}}}|^{\frac{(\gamma+2)\hat{\Delta}}{\Delta}+\hat{\gamma}}  {\cal{F}}_\phi   \left({ \frac{h}{r_{\rm{YL}}} }\right)
.
\label{c1}
\end{eqnarray}
Comparing Eq.(\ref{c1}) with the form (\ref{ct}), one obtains
$\alpha = 2 + \gamma - 2\Delta$ and $\hat{\alpha}= \hat{\gamma} + 2 \hat{\Delta}$.
From  Eqs.(\ref{Gr}) and (\ref{YLe}),  the former is the standard scaling law (\ref{Ru}).
From Eq.(\ref{NewLY}), the latter can be expressed as
a third relation between the logarithmic-correction exponents, namely
\begin{equation}
 \hat{\alpha} =  2 \hat{\beta} - \hat{\gamma}
.
\label{NewW2}
\end{equation}

Eqs.(\ref{e1}) and (\ref{c1}) may  be rewritten as
\begin{eqnarray}
  e_\infty(t,h) & = & h^{2-\frac{\gamma}{\Delta}-\frac{1}{\Delta}} |\ln{h}|^{\frac{(\gamma+1)\hat{\Delta}}{\Delta}+\hat{\gamma}}  {\cal{F}}^\prime_\phi   \left({ \frac{h}{r_{\rm{YL}}} }\right)
,
\label{e2}
\\
 c_\infty(t,h) & = & h^{2-\frac{\gamma}{\Delta}-\frac{2}{\Delta}} |\ln{h}|^{\frac{(\gamma+2)\hat{\Delta}}{\Delta}+\hat{\gamma}}  {\cal{F}}^\prime_\phi   \left({ \frac{h}{r_{\rm{YL}}} }\right)
.
\label{c2}
\end{eqnarray}
Letting $t \rightarrow 0$ so that $r_{\rm{YL}}(t) \rightarrow 0$, one finds
\begin{eqnarray}
  e_\infty(h) & = & h^{2-\frac{\gamma}{\Delta}-\frac{1}{\Delta}} |\ln{h}|^{\frac{(\gamma+1)\hat{\Delta}}{\Delta}+\hat{\gamma}}
,
\label{e3}
\\
 c_\infty(h) & = & h^{2-\frac{\gamma}{\Delta}-\frac{2}{\Delta}} |\ln{h}|^{\frac{(\gamma+2)\hat{\Delta}}{\Delta}+\hat{\gamma}}
.
\label{c3}
\end{eqnarray}
From the leading behaviour one recovers Eqs.(\ref{epsilon}) and (\ref{sigma}).
From the logarithmic corrections, one obtains the counterpart
scaling relations (\ref{SRlog5}) and (\ref{SRlog6}).

\subsection{Hyperscaling for Logarithms}\index{scaling relations!hyperscaling}
\label{HyperscalingL}

The logarithmic analogue of the hyperscaling relation (\ref{Jo}) has a rather different
status than the other critical exponents, in that it necessitates consideration of finite-size effects (this aspect will be further discussed below).
Consider, therefore, a system of finite volume $N=L^d$.\index{scaling!finite-size}
The finite-size scaling  (FSS) of Lee-Yang edge  is given by
\begin{equation}
 \frac{r_1(L)}{r_{\rm{YL}}(t)}
 =
 {\cal{F}}
 \left(
         \frac{\xi_L(0,0)}{\xi_\infty(t,0)}
 \right)
,
\label{modFSS}
\end{equation}
where $\xi_L(0)$ is the correlation length of the finite-size system at $t=0$. To allow for logarithmic corrections to this quantity too, we assume the form (\ref{corrL}).
For finite systems, the cumulative density of zeros is the
fractional total of zeros up to a given point \cite{JaKe01},
\begin{equation}
 G_L(r_j(L)) = \frac{2j-1}{2L^d}
.
\label{GL}
\end{equation}

Fixing the ratio $r/r_{\rm{YL}}(t)$ in (\ref{GG}), and using the scaling relations previously derived, and then taking the limit $t\rightarrow 0$, one arrives at an expression for the critical cumulative distribution function,
\begin{equation}
 G_\infty(r,0) \sim r^{\frac{2-\alpha}{\Delta}} |\ln{r}|^{\hat{\alpha}-\frac{(2-\alpha)}{\Delta}
\hat{\Delta}}
.
\label{GGG}
\end{equation}

At $t=0$ and for sufficiently large $L$, the FSS expression (\ref{GL}) must converge to the infinite-volume expression (\ref{GGG}).
Equating them gives the FSS of the first Lee-Yang zero at
criticality to be
\begin{equation}
 r_1(L) \sim L^{-\frac{d\Delta}{2-\alpha}}
 \left(
  \ln{L}
 \right)^{\hat{\Delta} - \frac{\Delta \hat{\alpha}}{2-\alpha}}
.
\label{Rand}
\end{equation}
Inserting (\ref{xit}), (\ref{edge}), (\ref{corrL})  and (\ref{Rand}) into (\ref{modFSS}) recovers (\ref{Jo}) and yields the logarithmic equivalent to the hyperscaling relation, namely
\begin{equation}
  \hat{\alpha} =  d \hat{q} -  d \hat{\nu}
,
\label{NewK}
\end{equation}
The logarithmic scaling relations (\ref{NewW1}) and (\ref{NewW2}) but not (\ref{NewK}) can be derived starting with a suitable modification to the phenomonological Widom  ansatz\cite{Ak01,Ke04}, namely
\begin{equation}
 f(t,h) = b^{-d}  f(b^{y_t}(\ln{b})^{\hat{y}_t}t,b^{y_h}(\ln{b})^{\hat{y}_h}h)
.
\label{Widomforlogs}
\end{equation}
Following the approach of Sec.\ref{Static}, this recovers the static scaling equations (\ref{Ru}), (\ref{Gr}), (\ref{epsilon}) and (\ref{sigma}) as well as their logarithmic counterparts (\ref{SRlog2}), (\ref{SRlog3}), (\ref{SRlog5}) and (\ref{SRlog6}) with
\begin{equation}
 \hat{y}_t = \frac{\hat{\beta}}{\beta} - \frac{\delta \hat{\delta}}{\beta(1+\delta)}
          = \frac{2 \hat{\beta} - \hat{\gamma}}{2 \beta + \gamma}
\label{delivery1}
\end{equation}
and
\begin{equation}
 \hat{y}_h = \frac{\delta \hat{\delta}}{1+\delta}
          = \frac{\gamma \hat{\beta} + \beta \hat{\gamma}}{2 \beta + \gamma}.
\label{delivery2}
\end{equation}
The logarithmic counterparts of Eqs.(\ref{alpha1}) and (\ref{delta1}) are then
\begin{equation}
 \hat{\alpha} = d \frac{\hat{y}_t}{y_t}, \quad
  \hat{\beta} = \beta \hat{y}_t + \hat{y}_h, \quad
 \hat{\gamma} = -\gamma \hat{y}_t + 2 \hat{y}_h
 \label{alpha1hat}
 \end{equation}
and
\begin{equation}
 \hat{\delta} = d \frac{\hat{y}_h}{y_h}, \quad
 \hat{\epsilon} = \hat{y}_t + \epsilon \hat{y}_h , \quad
 \hat{\alpha}_c = 2 \hat{y}_t - \alpha_c \hat{y}_h ,
\label{delta1hat}
\end{equation}
while
\begin{equation}
 \hat{\Delta} = \Delta \hat{y}_t - \hat{y}_h
.
\label{DDDhat}
\end{equation}

\subsection{Logarithmic Counterpart to Fisher's  Relation\index{scaling relations!Fisher}}
\label{LCFishers}

With multiplicative logarithmic corrections, the correlation function is given by Eq.(\ref{corrfun}).
Fixing the argument of the function $D$, we may rewrite this as
\begin{equation}
 {\cal{G}}_\infty (x,t,0)  \sim
 \xi_\infty(t,0)^{-(d-2+\eta)}(\ln{\xi_\infty(t,0)})^{\hat{\eta}}
 D\left( \frac{x}{\xi_\infty(t,0)} \right)
.
\label{corrfun1}
\end{equation}
Following Fisher \cite{Fi64sc}  and writing the susceptibility in terms of the correlation function (Eq.(\ref{Fishertrick})), one obtains
\begin{equation}
 \chi_\infty(t,0) \sim \xi_\infty(t,0)^{2-\eta} (\ln{\xi_\infty(t,0)})^{\hat{\eta}}
.
\label{bbq2}
\end{equation}
As in Sec.\ref{Fishers}, the leading power-laws in Eqs.(\ref{chit}) and (\ref{xit}) then deliver the scaling relation (\ref{Fi}).
Matching the logarithmic corrections exponents too yields
\begin{equation}
 \hat{\eta} =  \hat{\gamma} - \hat{\nu} (2 - \eta)
.
\label{red}
\end{equation}

This approach was used in Ref.[\refcite{SaSo97}] for the case of the $d=2$ four-state Potts model.\index{Potts model}
An alternative argument presented there (see Sec.\ref{Fishers}) assumes that the spins decorrelate  in the large-distance limit,
\begin{equation}
 {\cal{G}}_\infty (x,t) = \langle {\vec{s}}(0){\vec{s}}(x)\rangle
\rightarrow \langle {\vec{s}}(0)\rangle \langle {\vec{s}}(x)\rangle
= m_\infty^2(t,0).
\label{green3}
\end{equation}
Using Eqs.(\ref{xit}) and (\ref{corrfun1}) for the left hand side and
matching with Eq.(\ref{mt}) on the right, one again obtains
the standard scaling relation  (\ref{Fi}) from the leading exponents
 (see also Refs.\refcite{Ab67g,Su68}).
However, it is interesting to note that matching the logarithmic exponents gives, instead of  (\ref{red}), the relation
$
\hat{\eta} = 2\hat{\beta}
+ \hat{\nu} (d-2+\eta)
$.
From the logarithmic scaling relation (\ref{SRlog2}) this would mean
\begin{equation}
 \hat{\eta} = d \hat{q} + \hat{\gamma} - \hat{\nu} (2 - \eta)
,
\label{green2}
\end{equation}
which is, in general, {\emph{different} to Eq.(\ref{red}).

When $\hat{q}$ vanishes, as is the case in the
$d=2$, four-state Potts model \cite{SaSo97}, for example,
this is actually identical to Eq.(\ref{red}).
To decide between Eqs.(\ref{red}) and (\ref{green2}), we need a
model with non-zero $\hat{q}$ value. Models at the upper critical
dimension provide suitable determinators and one finds that
the Eq.(\ref{green2}) fails in these cases. Eq.(\ref{red}) holds in each case we have examined.

\subsection{Corrections to the Logarithmic Scaling Relations}
\label{CorrectScaling}

Although it holds in most models which manifest multiplicative logarithmic corrections, there is an immediate and obvious problem with scaling relation (\ref{NewK}) when it is confronted with the Ising model\index{Ising model} in two dimensions, which has been solved exactly.
There, the specific heat has a logarithmic divergence so that $\alpha = 0$ and $\hat{\alpha}=1$. There are, however, no logarithmic corrections in any of the other thermodynamic or correlation functions in this model, and, since $\hat{\nu}=\hat{q}=0$, Eq.(\ref{NewK}) fails immediately.
The relation (\ref{NewK}) also fails in the uncorrelated, quenched,  random disordered version of the Ising model in two dimensions, where ${\hat{q}} = 0$  \cite{Aade96,LaIg00},
$\hat{\alpha} =0 $, $\hat{\nu}   = 1/2  $  and \cite{SSLJ,DD,AnDo90WaSe90,KeJo06,GJannouncement,GJ84}
\begin{equation}
C_\infty(t,0) \sim \ln{|\ln{{t}}|}
.
\label{dL}
\end{equation}
This famous double logarithm in the specific heat of the diluted Ising model in two dimensions has been the source of great controversy throughout the years. This controversy has only recently been convincingly resolved, partly with the aid of logarithmic-correction theories\cite{KeRu08,GoKe09}, which should be able to account for them.
In this section, besides resolving the puzzle as to the value of $\hat{\alpha}$ in the pure and random Ising models in $d=2$ we will see how this double logarithm emerges quite naturally from the general scheme.

Since the problem is associated with the even sector of the model (namely with the $t$-dependency of the specific heat), one may argue
that a Lee-Yang analysis, which focuses on an odd (magnetic) scaling field, is not the best approach to fully extract the general relationship between the correction exponents appearing in Eq.(\ref{NewK}).
One may appeal to complex-temperature (Fisher) zeros\index{zeros!Fisher} for further insight, as they are appropriate to the even sector.
We will see that the puzzle is resolved after consideration of two special properties of the pure and random two-dimensional Ising models,
namely the vanishing of $\alpha$ and the angle at which their Fisher zeros impact onto the real axis.

An FSS theory\index{scaling!finite-size} for Fisher zeros is obtained \cite{IPZ} by writing the finite-size partition function in terms of the scaling ratio $\xi_L/\xi_\infty$
\begin{equation}
Z_L(t,0) = Q\left( \frac{\xi_L(0)}{\xi_\infty(t,0)} \right)
.
\label{Zzerot}
\end{equation}
This vanishes at a Fisher zero.
Labeling the $j^{\rm{th}}$ finite-size Fisher zero as $t_j(L)$, one therefore has
\begin{equation}
 \frac{\xi_L(0)}{\xi_\infty(t_j(L))} = Q_j^{-1}(0)
,
\end{equation}
where $Q_j^{-1}(0)$ is the $j^{\rm{th}}$ root of the function $Q$.
Using Eqs.(\ref{edge}) and (\ref{corrL}), this equation can be solved to give the FSS form of the $j$th zero,
\begin{equation}
 t_j(L) \sim L^{-\frac{1}{\nu}}
 (\ln{L})^{
            \frac{ \hat{\nu}-\hat{q} }{\nu}}
.
\label{A}
\end{equation}
So far, no assumptions other than the validity of FSS  have been involved.
Note that this is the same scaling form as  Eq.(\ref{SRlog8}), which is sensible, since the real part of the lowest zero is also a pseudocritical point.

The full expression for the scaling of the $j^{\rm{th}}$ Fisher zero\index{zeros!Fisher} is \cite{JaKe01,IPZ,GPS} a function of a fraction of the total number of
zeros $\cal{N}$. Since this is proportional to the lattice volume $N = L^d$, Eq.(\ref{A}) is more appropriately
written as
\begin{equation}
 t_j(L) \sim
 \left(\frac{j-1/2}{L^d}
 \right)^{\frac{1}{\nu d}}
 \left(
    \ln{\left( \frac{j-1/2}{L^d} \right)}
 \right)^\frac{\hat{\nu}-\hat{q}}{\nu}
 \exp{(i \phi_j(r_j(L)))}
,
\label{most}
\end{equation}
where $\phi_j(r_j(L))$ is the argument of the $j$th Fisher zero.

In all cases known from the literature, the Fisher zeros for isotropic models on homopolygonal lattices  fall on curves\index{zeros!locus}
in the complex plane. They impact onto the real axis along so-called singular lines \cite{Fi65,Ab1070,MaSh95}.
We also assume this scenario in the present case, and we denote the impact angle onto the real axis in the thermodynamic limit by  $\phi$.

Similar to the Lee-Yang case, we may write the finite-size partition function in terms of Fisher zeros,
\begin{equation}
 Z_L(t,0) \propto \prod_{j=1}^{\cal{N}} \left(t - t_j(L)\right)\left(t - t^*_j(L)\right)
,
\end{equation}
where we have been careful to identify $t_j(L)$ and $t^*_j(L)$ as complex conjugate pairs.
Provided that the
${\cal{M}} \propto {\cal{N}}$
zeros which dominate  scaling behavior in the vicinity of the critical point are described by the scaling form (\ref{most}),
and  differentiating appropriately, one finds the FSS for the  specific heat at $t=0$ to be
\begin{equation}
 C_L(0) \sim - L^{-d} {\rm{Re}}\sum_{j=1}^{\cal{M}} t_j^{-2}(L)
.
\label{Cc}
\end{equation}

We wish to compare this expression with that resulting from a direct application of the modified FSS hypothesis\index{scaling!finite-size} (\ref{modFSS}) to the  specific heat, which yields
\begin{equation}
 \frac{C_L(0)}{C_{\infty}(t,0)}
 =
 {\cal{F_C}}
 \left(
         \frac{\xi_L(0,0)}{\xi_\infty(t,0)}
 \right)
.
\label{modFSSC}
\end{equation}
Fixing the scaling ratio on the right hand side so that
$t \sim L^{-1/\nu}(\ln{L})^{(\hat{\nu}-\hat{q})/\nu}$,
one obtains from (\ref{ct}) the FSS behaviour
\begin{equation}
 C_L(0)
 \sim
 L^{\frac{\alpha}{\nu}}
 (\ln{L})^{\hat{\alpha} - \alpha \frac{\hat{\nu}-\hat{q}}{\nu}}
.
\label{B}
\end{equation}
We now match Eq.(\ref{Cc}) to Eq.(\ref{B}).

In the case where $\alpha \ne 0$, Eq. (\ref{Cc}) gives
\begin{equation}
 C_L(0) \sim L^{\frac{\alpha}{\nu}}  \left(
    \ln{L}
 \right)^{-2\frac{\hat{\nu}-\hat{q}}{\nu}}
,
\label{D}
\end{equation}
and comparison with Eq.({\ref{B}) leads to the recovery of the previously derived logarithmic scaling relation~(\ref{NewK}).

If, however, $\alpha = 0$,  the FSS expression (\ref{Cc})
for the specific heat becomes
\begin{equation}
 C_L(0) \sim
 \sum_{j=1}^{\cal{M}}{
 \frac{\cos{(2\phi_j(L))}}{j-1/2}
 \left(
    \ln{\left( \frac{j-1/2}{L^d} \right)}
 \right)^{-2 \frac{\hat{\nu}-\hat{q}}{\nu}}
}
.
\label{enrage}
\end{equation}
For large $L$ and close enough to the transition point that
$\phi_j(r_j(L)) \simeq \phi$, the cosine term becomes a non-vanishing
constant if  $\phi \ne \pi/4$.
This is what happens in the  pure Ising model in $d=2$ dimensions,
where $\phi = \pi/2$  is assured by duality on the square lattice \cite{Fi65}
as well as for its symmetric random-bond counterpart \cite{Fi78}. Universality of $\phi$ ensures this happens for
other lattice configurations too and continuity arguments
lead one to also expect $\phi \ne \pi/4$
in the general random-bond and random-site two-dimensional  Ising case.\index{Ising model}

In these cases, application of the Euler-Maclaurin formula gives that
the leading FSS  behavior when $\alpha=0$ is
\begin{equation}
 C_L(0)
 \sim
 \left\{
        \begin{array}{ll}
                         (\ln{L})^{
                                   1-2 \frac{ \hat{\nu}-\hat{q} }{ \nu }
                                  } & \mbox{
                                            if $2(\hat{\nu}-\hat{q}) \ne \nu$
                                           } \\
                        \ln{\ln{L}} & \mbox{
                                            if $2(\hat{\nu}-\hat{q}) = \nu$
                                            } .
        \end{array}
 \right.
\label{doublelog}
\end{equation}

In the thermodyamic limit, one may simply replace $L$ and $C_L(0)$ by ${t}$ and $C_\infty(t,0)$  in Eq.(\ref{doublelog}), respectively.
Then, comparing  Eq.(\ref{B}) with Eq.(\ref{doublelog}) and using standard hyperscaling (\ref{Jo}), one finds
\begin{equation}
 \hat{\alpha}
 =
 1 + d \hat{q} - d\hat{\nu}
.
\label{NewK2}
\end{equation}
This expression replaces Eq.(\ref{NewK}) when the model has
both $\alpha = 0$ and $\phi \ne \pi/4$.
Thus the logarithmic scaling relation (\ref{SRlog1}) is established.
Furthermore, we prefer to write Eq.(\ref{SRlog2}) in terms of $\hat{q}$ and $\hat{\nu}$ rather than in terms of $\hat{\alpha}$ as in Eq.(\ref{NewW2}) because, unlike $\hat{\alpha}$,  $\hat{q}$ and $\hat{\nu}$  the $\hat{q}-\hat{\nu}$ combination does not exhibit the subtleties discussed here.

We can perform similar considerations for the Lee-Yang\index{zeros!Lee-Yang} zeros, replacing
Eq.(\ref{Zzerot}) by
\begin{equation}
Z_L(h) = Q\left( \frac{\xi_L(0)}{\xi_\infty(0,h)} \right)
.
\label{Zzeroh}
\end{equation}
We find
\begin{equation}
 \hat{\delta} = d (\hat{q}-\hat{\nu}_c)
 ,
 \label{derss}
\end{equation}
except when $\gamma = 0$ and the impact angle is not $\pi/4$,\index{impact angle}
in which case an extra logarithm appears and Eq.(\ref{derss}) becomes
\begin{equation}
 \hat{\delta} = 1+ d (\hat{q}-\hat{\nu}_c)
 .
 \label{derss2}
\end{equation}
That the impact angle is $\pi/2$ is guaranteed by the Lee-Yang theorem,
but we know of no cases with $\gamma = 0$, so we omit Eq.(\ref{derss}) from Eq.(\ref{SRlogX}).

To summarize so far, we have explained where the standard scaling relations come from using the standard approaches of Widom scaling and the Kadanoff block-spin RG. These standard scaling laws for the leading critical exponents are well established in the literature.
Analogous relations for the logarithmic corrections
(\ref{SRlog1})--(\ref{SRlog7}) have also been derived using a self-consistency approach.
Next we confront the logarithmic scaling relations with results from the
literature in  a variety of models. Where predictions for, or measurements of, these logarithmic corrections already exist in the literature, we can test the new scaling relations. Indeed, we will show that they are upheld (barring at least 3 exceptional circumstances which require further research). Where there are gaps in the literature regarding the values of the logarithmic corrections, we can use the scaling relations to make predictions for them. These new predictions need to be independently tested in future research.

\subsection{The Logarithmic Shift Exponent}
\label{shiftseclog}

Finally we mention that the finite-size scaling  of the pseudocritical point\index{shift exponent}\index{pseudocritical point}  may be determined using the Widom static\index{scaling!Widom}\index{scaling!static} hypothesis (\ref{Widomforlogs}) modified to include logarithms.
In a similar manner to Sec.\ref{shiftsec}, one finds that the specific heat peaks scales as Eq.(\ref{shift2}) with
\begin{equation}
 \hat{\lambda}_{\rm{shift}}
 =
 -\hat{y}_t
 =
 \frac{\hat{\nu}-\hat{q}}{\nu}
.
\label{SRlogY}
\end{equation}
This is the final scaling relation for logarithmic corrections Eq.(\ref{SRlog8}).

\section{Fisher Renormalization for Logarithmic Corrections}
\label{Fren}

Fisher renormalization\index{renormalization!Fisher} concerns systems under some form of\index{constraint} constraint. For such systems, the critical exponents take values which differ from their ``ideal'' counterparts. The systems we have been dealing with so far are ``ideal'', in the sense that they are not subject to constraints of this type. Typically, the theoretical power-law divergence of the specific heat in an ideal system, for example, is replaced by a finite cusp in a ``real'' experimental realization.
Fisher \cite{Fi68} explained this as being due to the effect of hidden variables and established elegant relations between the ideal exponents and their constrained counterparts.

Phase transitions which exhibit Fisher renormalization include those in
constrained magnetic and fluid systems (e.g., with fixed levels of impurities),  
the superfluid $\lambda$  transition  in $^3$He-$^4$He mixtures in confined films \cite{3He4He}, 
the order-disorder transition in compressible ammonium chloride \cite{ImEn73,Na4Cl}, 
the critical behaviour at nematic-smectic-A transitions in liquid-crystal mixtures \cite{dePr95} and in emulsions \cite{paint}.

When multiplicative logarithmic corrections are present in the ideal system, these migrate to the Fisher-renormalized system in a non-trivial manner which has recently been determined \cite{ourFren}.
Here, we summarize the Fisher renormalization for the logarithmic exponents. We later use this scheme to deduce the scaling behaviour of lattice animals at their upper critical dimension.

For a system under constraint, Fisher established that if the specific-heat exponent for the ideal system $\alpha$
is positive, it is altered in the constrained system.
The magnetization, susceptibility and  correlation-length
critical exponents are also changed. If the subscript ``$X$'' represents the real or constrained system, its critical exponents are related to the ideal ones by the transformations
\begin{equation}
\alpha_X  = \frac{-\alpha}{1-\alpha},
\quad
\beta_X = \frac{\beta}{1-\alpha},
\quad
\gamma_X = \frac{\gamma}{1-\alpha},
\quad
\nu_X = \frac{\nu}{1-\alpha}.
\end{equation}
The exponent $\delta$ and the anomalous dimension\index{dimension!anomalous} $\eta$ are not renormalized:
\begin{equation}
\delta_X = \delta  ,
 \quad
 {\mbox{and }}
 \quad
\eta_X = \eta.
\label{Fre}
\end{equation}
Expressing Eqs.(\ref{ytyh}) in terms of the above exponents, and Fisher renomalizing, gives the RG eigenvalues for the constrained system as
\begin{equation}
 {y_t}_X = (1-\alpha) y_t, \quad {\mbox{and}} \quad
 {y_h}_X = y_h.
\end{equation}
One may then use the scaling relations (\ref{X}),  (\ref{YLe}),
(\ref{shiftexp}) and (\ref{delta1}) to determine the Fisher-renormalization formulae
\begin{equation}
\Delta_X = \frac{\Delta}{1-\alpha}, \quad
{\nu_c}_X = \nu_c, \quad
\epsilon_X = \epsilon + \frac{\alpha}{\Delta}, \quad
{\alpha_c}_X = \alpha_c - 2 \frac{\alpha}{\Delta}.
\end{equation}
as well as
\begin{equation}
{\lambda}_{\rm{shift} X} = d - \lambda_{\rm{shift}}.
\end{equation}

These formulae have two attractive properties:
\begin{itemize}
\item If the ideal exponents obey the standard scaling relations
then the Fisher renormalized exponents  do likewise.
\item Fisher renormalization is involutory\index{involutory} in the sense that  Fisher renormalization of the constrained exponents returns the ideal ones. This means that two successive applications of the transformation gives the identity.
\end{itemize}

Fisher renormalization applied to the most commonly encountered logarithmic correction exponents results in \cite{ourFren}
\begin{eqnarray}
 \hat{\alpha}_X & = & -\frac{\hat{\alpha}}{1-\alpha}
,
\label{Fa}
\\
 \hat{\beta}_X & = & \hat{\beta} -\frac{\beta \hat{\alpha}}{1-\alpha}
,
\label{Fb}
\\
 \hat{\gamma}_X & = & \hat{\gamma} +\frac{\gamma \hat{\alpha}}{1-\alpha}
,
\label{Fg}
\\
\hat{\nu}_X  & = &\hat{\nu} +\frac{\nu \hat{\alpha}}{1-\alpha}
.
\label{Fn}
\end{eqnarray}
As for the leading indices, no Fisher renormalization occurs for the logarithmic-correction exponents for the in-field magnetization
or for the correlation function, since  it is defined exactly at the critical point. Similarly $\hat{q}$ is unchanged.
\begin{equation}
 \hat{\delta}_X  =  \hat{\delta}, \quad
 \hat{\eta}_X  = \hat{\eta} , \quad
  \hat{q}_X  =  \hat{q} .
\label{Fq}
\end{equation}
Each of these obey the scaling relations for logarithmic corrections,  and
Eqs.(\ref{delivery1}) and (\ref{delivery2}) give
\begin{eqnarray}
 \hat{y}_{t X} & = & (1-\alpha ) \hat{y}_t - \hat{\alpha}
 ,
 \label{delivery1F} \\
 \hat{y}_{h X} & = & \hat{y}_h,
 \label{delivery2F}
\end{eqnarray}
while
Eqs.(\ref{SRlog5})--(\ref{SRlog8}) also lead to
\begin{eqnarray}
 \hat{\epsilon}_X & = & \hat{\epsilon} - \hat{\alpha} - \frac{\alpha \hat{\Delta}}{\Delta},
 \label{Fepsilonhat} \\
 {\hat{\alpha}}_{cX} & = & \hat{\alpha}_c -2\left({ \hat{\alpha} + \frac{\alpha \hat{\Delta}}{\Delta} }\right),
 \label{FalphahatC}  \\
{\hat{\nu}}_{cX} & = & {\hat{\nu}}_{c} ,
 \label{FnuChat} \\
\hat{\Delta}_X & = & \hat{\Delta} - \frac{\Delta \hat{\alpha}}{1-\alpha},
 \label{FDeltahat}
 \\
{\hat{\lambda}_{\rm{shift} X}} & = & (1-\alpha) \hat{\lambda}_{\rm{shift}} + \hat{\alpha}.
 \label{Flambdahat}
 \end{eqnarray}
Fisher renormalization is also involutory at the logarithmic level: re-renormalizing the Fisher-renormalized logarithmic exponents returns   their ideal counterpart. In other words, Fisher renormalization is  its own inverse at the logarithmic as well as leading level.

Fisher renormalization at the logarithmic level has been tested in lattice animals and the Lee-Yang\index{Lee-Yang!problem} problem at their upper critical dimensions, which also led to new predictions for logarithmic corrections are made. These are discussed in Sec.\ref{LatticeAnimals} below.

\section{Logarithmic Correction Exponents for Various Models}
\label{Models}

In this section we confront the logarithmic scaling relations (\ref{SRlog1})--(\ref{SRlog7}) with the literature for various models on a case-by-case basis.

\subsection{$q=4$ $d=2$ Potts Model}\index{Potts model}
\label{Potts}

The leading critical exponents for the  $4$-state Potts model in $d=2$ dimensions are \cite{leadP,ShchurBerche}
\begin{eqnarray}
\alpha &= & \frac{2}{3}, \quad
\beta=\frac{1}{12}, \quad
\gamma=\frac{7}{6}, \quad
\delta=15         , \quad
\nu=\frac{2}{3}   , \nonumber \\
\eta & = & \frac{1}{4}, \quad
\epsilon= \frac{4}{15}       , \quad
\alpha_c= \frac{8}{15}       , \quad
\nu_c= \frac{8}{15}          , \quad
\Delta=\frac{5}{4}.
\end{eqnarray}
The hitherto-known logarithmic-correction exponents are \cite{NaSc80CaNa80,SaSo97,ShchurBerche}
\begin{eqnarray}
\hat{\alpha}=-1,    \quad
\hat{\beta}=-\frac{1}{8}, \quad
\hat{\gamma} = \frac{3}{4}, \quad
\hat{\delta}=-\frac{1}{15}, \quad
\hat{\nu} = \frac{1}{2},    \quad
\hat{\eta} = -\frac{1}{8}.  \quad
\end{eqnarray}
FSS of the thermodynamic functions \cite{SaSo97,BlEm81} means that
\begin{equation}
\hat{q}=0.
\end{equation}
The correction relations (\ref{SRlog1})--(\ref{SRlog4}) therefore hold,
while (\ref{YLe}) gives $\Delta=5/4$ for the leading scaling of the Lee-Yang edge.\index{Lee-Yang!edge}
We can also use the remaining static scaling laws (\ref{SRlog5})--(\ref{SRlogX})  to  predict
\begin{equation}
\hat{\epsilon}= -\frac{23}{30},   \quad
\hat{\alpha}_c= - \frac{22}{15},   \quad
\hat{\nu}_c = \frac{1}{30},  \quad
\hat{\Delta} = -\frac{7}{8}.
\end{equation}

\subsection{$O(N)$  $\phi^4_d$ Theory}\index{$\phi^4$-theory}
\label{phi4}

The upper critical dimension for  $O(N)$-symmetric $\phi^4_d$ theories -- of crucial importance to the Higgs sector of the standard model --
is $d_c=4$. Here hyperscaling fails and the leading critical exponents take on their mean-field values which are
\begin{eqnarray}
\alpha & = & 0, \quad
\beta=\frac{1}{2}, \quad
\gamma=1, \quad
\delta=3,\quad
\nu=\frac{1}{2},  \nonumber  \\
\eta & = & 0,\quad
\epsilon= \frac{2}{3}, \quad
\alpha_c= 0,\quad
\nu_c= \frac{1}{3}, \quad
\Delta= \frac{3}{2}.\quad
\end{eqnarray}
The RG predictions for the corrections are already known to be \cite{Ke04,Br82,BLZLuWe89,KeLa91}
\begin{eqnarray}
\hat{\alpha} & = & \frac{4-N}{N+8}, \quad
\hat{\beta}=\frac{3}{N+8}, \quad
\hat{\gamma} = \frac{N+2}{N+8}, \quad
\hat{\delta}=\frac{1}{3}, \quad  \\
\hat{\nu} & = & \frac{N+2}{2(N+8)}, \quad
\hat{\eta} = 0, \quad
\hat{\Delta} = \frac{1-N}{N+8}, \quad
\hat{q}=\frac{1}{4}. \quad
\end{eqnarray}
The correction relations (\ref{SRlog1})--(\ref{SRlog4}) and (\ref{SRlog7}) therefore hold and from formulae (\ref{SRlog6})--(\ref{SRlogX}) we predict
\begin{equation}
\hat{\epsilon}=\frac{10-N}{3(N+8)}, \quad
\hat{\alpha}_c=\frac{4-N}{N+8}, \quad
\hat{\nu}_c = \frac{1}{6}.
\end{equation}
In the four-dimensional Ising  case it is known \cite{IPZ,GPS,KeLa91}
that the impact angle for Fisher zeros is $\phi=\pi/4$.
Assuming the same for its $O(N)$ generalization, means that
Eq.(\ref{NewK2}) does not follow from Eq.(\ref{enrage}) and
Eq.(\ref{NewK}) remains valid there instead.
The same is expected to be true for $O(N)$ theories with long-range interactions, which we next address.

\subsection{Long-Range $O(N)$  $\phi^4_d$ Theory}\index{long-range systems}\index{$\phi^4$-theory!long-range}
\label{LRON}

The introduction of long-range interactions alters the universality class of $O(N)$ spin models. If the interactions decay as $x^{-(d+\sigma)}$, the upper critical dimension becomes $d_c=2\sigma$.
The critical exponents for the $N$-component long-range system are \cite{FiMa72,LuBl97}
\begin{eqnarray}
\alpha=0,\quad
\beta=\frac{1}{2},\quad
\gamma=1,\quad
\delta=3,\quad
\nu=\frac{1}{\sigma},\quad   \\
\eta = 2-\sigma,\quad
\epsilon= \frac{2}{3},\quad
\alpha_c= 0,\quad
{\nu}_c= \frac{1}{3},\quad
\Delta= \frac{3}{2},
\end{eqnarray}
which obey the leading scaling relations.
The Privman-Fisher form for the free energy \cite{LuBl97} gives the RG values for the logarithmic corrections to be
\begin{eqnarray}
\hat{\alpha} & = & \frac{4-N}{N+8},\quad
\hat{\beta}=\frac{3}{N+8},\quad
\hat{\gamma} = \frac{N+2}{N+8}, \nonumber \\
\hat{\delta} & = & \frac{1}{3},\quad
\hat{\nu} = \frac{N+2}{\sigma(N+8)},\quad
\hat{\eta} = 0.\quad
\end{eqnarray}
The correction relations (\ref{SRlog2}) -- (\ref{SRlog4}) are obeyed, and the remaining relations (\ref{SRlog1}) and (\ref{SRlog5})--(\ref{SRlog7})  predict
\begin{equation}
\hat{\epsilon}=\frac{10-N}{3(N+8)},\quad
\hat{\alpha}_c= \frac{4-N}{N+8} ,\quad
\hat{\nu}_c = \frac{6-\sigma}{12 \sigma},\quad
\hat{\Delta} = \frac{1-N}{N+8},\quad
\hat{q}= \frac{1}{2\sigma}. \\
\end{equation}
The prediction  $\hat{q}=1/2\sigma$ recovers the known value\cite{Br82} $\hat{q}=1/4$ for $O(N)$ $\phi^4_4$ theory  in the $\sigma = 2$ case.
It also leads to agreement with long-range Ising FSS in two dimensions
when $\sigma = 1$  \cite{GrHu04}.
The remaining predictions for long-range systems have yet to be verified.

\subsection{Spin Glasses in 6 Dimensions}\index{spin glass}\index{percolation}\index{Lee-Yang!edge}\index{lattice animals}\index{$\phi^3$-theory}
\label{SpinGlass}

Spin glasses, percolation, the  Lee-Yang edge, and  lattice animals problems are each  related to $\phi^3$ field theory.
The leading exponents are \cite{EdAn75HaLu76,Pottslogs,Ru98}
\begin{equation}
\alpha  =  -1, \quad
\beta = 1,     \quad
\gamma = 1,    \quad
\delta = 2,    \quad
\nu=\frac{1}{2}, \quad
\eta= 0,
\label{RLMF}
\end{equation}
and obey the standard scaling relations provided
\begin{equation}
\Delta=2 , \quad
\epsilon = 1 , \quad
\alpha_c = -\frac{1}{2} , \quad
\nu_c = \frac{1}{4},
\label{predic}
\end{equation}
predictions which still need numerical verification.
Ruiz-Lorenzo used RG methods to derive the critical scaling exponents
of the correlation length, susceptibility and specific heat for these  models at their upper critical dimensions as \cite{Ru98}
\begin{equation}
\hat{\alpha}  = \frac{2(2b-3a)}{4b-a}, \,
\hat{\gamma} = \frac{2a}{4b-a}, \,
\hat{\nu}   =\frac{5a}{6(4b-a)},
\label{RLcorr}
\end{equation}
where the values of $(a,b)$ depend upon which problem one is considering.

In the $m$-component  spin-glass case, the upper critical dimension is $d_c=6$ and values of $(a,b)$ are $(-4m,1-3m)$.
There, the leading exponents are given by Eq.(\ref{RLMF}) while Eq.(\ref{RLcorr}) gives the logarithmic correction exponents as
\begin{equation}
\hat{\alpha}  =  -\frac{3m+1}{2m-1}, \quad
\hat{\gamma} = \frac{2m}{2m-1}, \quad
\hat{\nu}   =\frac{5m}{6(2m-1)}.
\end{equation}
These satisfy the scaling relations
(\ref{SRlog1}) -- (\ref{SRlog7}) provided that
\begin{eqnarray}
\hat{\beta}    & = &             \frac{1+m}{2(1-2m)},\quad
\hat{\delta}     =               \frac{1-3m}{4(1-2m)},\quad
\hat{\eta}       =               \frac{m}{3(2m-1)} ,\quad
\hat{\epsilon}   =   \frac{1}{2} \frac{1+2m}{1-2m} , \nonumber \\
\hat{\alpha}_c & = & \frac{1}{4} \frac{3+7m}{1-2m} ,\quad
\hat{\nu}_c      =   \frac{1}{24} \frac{5m-3}{2m-1} ,\quad
\hat{\Delta}     =   \frac{1+5m}{2(1-2m)},\quad
\hat{q}          =   \frac{1}{6} .
\end{eqnarray}
Independent numerical investigations of these logarithmic corrections are required. In particular, Ruiz-Lorenzo's prediction for the finite-size correlation-length correction exponent is $\hat{q}=1/3$ \cite{Ru98}.
This is the first instance where we encounter a clash with results in the literature regarding the exponent $\hat{q}$, the other two being in the percolation and the  Lee-Yang/lattice-animal problems. These cases are discussed next.

\subsection{Percolation in 6 Dimensions}
\label{Percolation6}

The percolation problem at its upper critical dimension of $d_c=6$ has \cite{EdAn75HaLu76,Pottslogs} the same leading, mean-field  critical exponents as in Eqs.(\ref{RLMF}) and ({\ref{predic}).
With $(a,b) = (-1,-2)$, in the percolation case
the following correction exponents are known \cite{Pottslogs,Ru98,StJa03}
\begin{equation}
\hat{\alpha}  =  \frac{2}{7} , \quad
\hat{\beta}   =  \frac{2}{7} , \quad
\hat{\gamma}  =  \frac{2}{7} , \quad
\hat{\delta}  =  \frac{2}{7}  , \quad
\hat{\nu}     =  \frac{5}{42}  , \quad
\hat{\eta}    =  \frac{1}{21}  .
\end{equation}
The scaling relations for logarithmic corrections now allow the prediction
\begin{equation}
\hat{\epsilon}  = \frac{2}{7}  , \quad
\hat{\alpha}_c  = \frac{2}{7} , \quad
\hat{\nu}_c    =  \frac{5}{42} , \quad
\hat{\Delta}   =  0  , \quad
\hat{q}     = \frac{1}{6}.
\end{equation}
While other works \cite{FoAh04} contain an implicit assumption that $\hat{q}=0$, Ruiz-Lorenzo's prediction\cite{Ru98} for this quantity is  $\hat{q}=1/3$. Again, further investigations are needed.

\subsection{Lee-Yang Problem in 6 Dimensions and Lattice Animals in 8 Dimensions}
\label{LatticeAnimals}

The lattice-animal problem \cite{LuIs78,percolation} is closely linked to the  Lee-Yang edge problem \cite{Fi78}. The former has
upper critical dimensionality $8$, while for the latter it is $6$.
The mean-field values of the critical exponents for the  Lee-Yang edge problem are again given by Eqs.(\ref{RLMF}) and (\ref{predic}), while the values $(a,b) = (-1,-1)$ for Lee-Yang singularities \cite{Ru98} lead to
\begin{equation}
\hat{\alpha}  =  -\frac{2}{3} , \quad
\hat{\gamma}  =  \frac{2}{3}  , \quad
\hat{\nu}     =  \frac{5}{18}  .
\end{equation}
The scaling relations for logarithmic corrections then predict
\begin{eqnarray}
\hat{\beta}           & = &        0                , \quad
\hat{\delta}            =          \frac{1}{3}      , \quad
\hat{\eta}              =          \frac{1}{9}      , \quad
\hat{\epsilon}  =    0     ,  \nonumber \\
\hat{\alpha}_c &= &  -\frac{1}{3}   , \quad
\hat{\nu}_c     =  \frac{1}{9}              , \quad
\hat{\Delta}            =         -\frac{2}{3}      , \quad
\hat{q}                 =         \frac{1}{6}       .
\end{eqnarray}
Fisher-renormalizing both the mean-field leading critical exponents and the logarithmic corrections, one obtains
(omiting the subscript $X$)
\begin{eqnarray}
\alpha &=&  \frac{1}{2}   , \quad
\beta   =   \frac{1}{2}   , \quad
\gamma  =   \frac{1}{2}   , \quad
\delta  =   2             , \quad
\nu     =   \frac{1}{4}   ,  \nonumber \\
\eta    &=&  0              , \quad
\Delta  =  1     , \quad
\epsilon   =   \frac{1}{2}   , \quad
\alpha_c   =   -\frac{1}{2} , \quad
\nu_c      =   \frac{1}{2} .
\end{eqnarray}
and
\begin{eqnarray}
\hat{\alpha} &=&  \frac{1}{3}   , \quad
\hat{\beta}   =   \frac{1}{3}   , \quad
\hat{\gamma}  =   \frac{1}{3}   , \quad
\hat{\delta}  =   \frac{1}{3}   , \quad
\hat{\nu}     =   \frac{1}{9}   , \nonumber \\
\hat{\eta}   &=&  \frac{1}{9}   , \quad
\hat{\Delta}   =    0               , \quad
\hat{\epsilon}   =   \frac{1}{3}    , \quad
\hat{\alpha}_c   =   \frac{1}{3}    , \quad
\hat{\nu}_c      =   \frac{1}{9}    .
\end{eqnarray}
together with
\begin{equation}
 \hat{q} = \frac{1}{6}.
\end{equation}
These deliver our theoretical predictions for the lattice-animal problem at its upper critical dimensionality $d=8$.

A lattice animal is a cluster of connected sites on a regular lattice.
Variants include clusters of connected bonds as well as weakly embedded and strongly embedded trees.
It is believed that these  models belong to the same universality class.
The enumeration of lattice animals  is a combinatorial problem of interest to mathematicians.
In physics, they are closely linked to  percolation and clustering in spin models.
In chemistry they form a basis for models of  branched polymers in good solvents.
Lattice animals linked by translations are considered as belonging to the same equivalence class, and as such are considered to
be essentially the same. 
One is interested in  $Z_N$, the number of animals containing $N$ sites, and the radius of gyration $R_N$, which is the average distance of  occupied sites
to the centre of mass of the  animal.
Allowing for logarithmic corrections, these behave as \cite{LuIs79}
\begin{eqnarray}
Z_N & \sim & \mu^{N}N^{-\theta} (\ln{N})^{\hat{\theta}},
\label{1}\\
R_N & \sim & N^{\nu} (\ln{N})^{\hat{\nu}}.
\label{2}
\end{eqnarray}
The exponent $\theta$ may be identified with $3-\alpha$, which, from the above Fisher-renormalized values is $5/2$.
There is a famous scaling relation due to Parisi and Sourlas\index{scaling relations!Parisi-Sourlas} which predicts that $\theta$ and $\nu$ are  related by
\cite{PaSo81}
\begin{equation}
 \theta = (d-2)\nu +1 .
 \label{PS}
\end{equation}
This is essentially hyperscaling\index{scaling relations!hyperscaling} (\ref{Jo}) with the $d$ dimensionally reduced by $2$ and has been numerically verified in dimensions $d=2$ to $d=9$ \cite{HsNa05}.
Identifying $\hat{\theta} = \hat{\alpha}$ The scaling relation (\ref{SRlog1}) may be written
\begin{equation}
 \hat{\theta} = (d-2)(\hat{q}-\hat{\nu}) = 6(\hat{q}-\hat{\nu}),
\label{Oureq16}
\end{equation}
having reduced the dimensionality term from $d=d_c=8$ to $6$, appropriately.
This is the logarithmic counterpart to the Parisi-Sourlas relation (\ref{PS}).
The above values $\hat{\alpha} = 1/3$, $\hat{q}=1/6$, $\hat{\nu}=1/9$ satisfy Eq.(\ref{Oureq16}) by construction. An alternative set of values in the literature \cite{Ru98} is $\hat{\alpha} = 1/3$, $\hat{q}=1/3$, $\hat{\nu}=-1/72$. This set comes directly from an RG-based calculation and does not satisfy the new scaling relation. Moreover, a recent numerical study \cite{voFo11}, though not conclusive,  indicates that the set of exponents developed here is more likely to be the correct one. Clearly more research is needed to explain the disparity of this set with the RG.

\subsection{Ising Model in 2 Dimensions}\index{Ising model}
\label{IM2}

The Ising model two dimensions has critical exponents
\begin{eqnarray}
{\alpha}  & = &  0            , \quad
{\beta}   =   \frac{1}{8}  , \quad
{\gamma}  =   \frac{7}{4}  , \quad
{\delta}  =   15           , \quad
{\nu}     =   1            , \nonumber \\
\hat{\eta}    & = &  \frac{1}{4}   , \quad
{\Delta}   =   0      , \quad
{\epsilon}   = 0      , \quad
{\alpha}_c   = 0      , \quad
{\nu}_c      = 0      .
\label{IM2exp}
\end{eqnarray}
The specific heat has a logarithmic divergence with temperature in this model so that $\hat{\alpha}=1$. However, this well-studied model is unusual in that there are no other logarithmic divergences and all the remaining hatted exponents vanish. The subtle nature of the emergence of this unusual behaviour was explained in Sec.\ref{CorrectScaling}: a logarithm arises from summing over the Fisher-zero indices, a summation which is necessary because these zeros approach the real-temperature axis at an angle other than $\pi/4$. In fact the angle of approach is $\pi/4$. This is because the self-dual nature of the two-dimensional Ising model assures a symmetry between the high- and low- temperature sectors, which demands that the impact angle $\phi = \pi/2$.

\subsection{Quenched-Disordered Ising Model in 2 Dimensions}\index{Ising model}
\label{IM2dis}

The uncorrelated, quenched, random removal of sites or the randomisation of bond strengths on a lattice is expected to immitate the presence of impurities in real physical systems.
We usually appeal to the Harris\index{Harris} criterion\cite{Ha74} to answer the question of how such randomisation affects the critical exponents \cite{Ha74}.
If $\alpha > 0$ in the pure system, quenched disorder is deemed relevant and the critical exponents change as the disorder is increased. On the other hand, if $\alpha < 0$ in the pure model, then disorder of this type does not change the critical behaviour and the exponents are unaltered.
In the case of $\alpha=0$, which as we have seen describes the Ising model in two dimensions, the Harris criterion does not  provide a clear  answer.

For this reason, the bond- and site-diluted Ising models in two dimensions have been controversial over the years.
The notion that some of the leading critical exponents change as the lattice structure is randomised, but where combinations which appear in terms of the correlation length such as $\beta/\nu$ and $\gamma/\nu$ are unchanged, became known as the {\emph{weak universality hypothesis}\/}.

Gradually this gave way to the {\emph{strong universality hypothesis}} \cite{KeRu08}, which is now mostly believed to hold, although agreement is not universal \cite{GoKe09}. This predicts that the two-dimensional diluted models  have the same leading critical exponents as in the pure case, but that there are multiplicative logarithmic corrections \cite{GJ83,SSLJ}
\begin{equation}
\hat{\alpha} =0                 , \quad
\hat{\beta}  =    -\frac{1}{16} , \quad
\hat{\gamma}  =   \frac{7}{8}   , \quad
\hat{\delta}  =   0             , \quad
\hat{\nu}     =   \frac{1}{2}   , \quad
\hat{\eta}    =   0             .
\end{equation}
With $\hat{q}=0$ \cite{Aade96},
these correction exponents obey the scaling relations for logarithmic corrections (\ref{SRlog1})--(\ref{SRlog4}).
The remaining scaling relations (\ref{SRlog5})--(\ref{SRlog7}) predict
\begin{equation}
\hat{\Delta}    =   -\frac{15}{16}   , \quad
\hat{\epsilon}  =   -\frac{1}{2}  , \quad
\hat{\alpha}_c  =   -1    , \quad
\hat{\nu}_c     = 0          .
\end{equation}
Moreover Eq.(\ref{doublelog}) means that there is a double logarithm in the specific heat \cite{SSLJ,DD,AnDo90WaSe90,KeJo06,GJannouncement,GJ84}
 \begin{equation}
 c_\infty(t,0) \sim \ln{|\ln{{t}}|}
.
\label{loglog}
\end{equation}
The power of the scaling relations for logarithmic corrections is well illustrated in this model as Eq.(\ref{SRlog1}) connects the hitherto most elusive and controversial quantity $\hat{\alpha}$ directly to other exponents, which are more clearly established.
For a review of the two-dimensional disordered Ising model, see Ref.[\refcite{GoKe09}].

\subsection{Ashkin-Teller Model in 2 Dimensions}\index{Ashkin-Teller Model}
\label{ATM2}

The new relation (\ref{red}) also holds in the $N$-colour Ashkin-Teller model
which consists of $N$ coupled Ising models. the leading exponents are the same as Eq.(\ref{IM2exp}) and \cite{GrWi81,SSLJ}
\begin{equation}
\hat{\alpha} = -\frac{N}{N-2}, \,
\hat{\beta}  = -\frac{n-1}{8(n-2)}, \,
\hat{\gamma} = \frac{7(N-1)}{4(N-2)}, \,
\hat{\delta} = 0, \,
\hat{\nu}   = \frac{N-1}{N-2}, \,
\hat{\eta}   = 0.
\end{equation}
If $\hat{q}=0$, these values also support the scaling relations (\ref{SRlog1})--(\ref{SRlog4}). Eqs.(\ref{SRlog5})--(\ref{SRlog7}) further lead to the predictions
\begin{equation}
\hat{\Delta}    =  -\frac{15}{8} \frac{n-1}{n-2}, \quad
\hat{\epsilon}  =  - \frac{n-1}{n-2} , \quad
\hat{\alpha}_c  =  -2 \frac{n-1}{n-2} , \quad
\hat{\nu}_c     =  0 .
\end{equation}

\subsection{Spin Models on Networks}\index{networks}
\label{Networks}

Here we consider another set of phase transitions which exhibit multiplicative logarithmic corrections to scaling, namely  spin systems on scale free networks.
There are both academic and practical motivations for studying critical phenomena on complex networks \cite{networks_rev}.
Phenomena such as opinion formation in social networks \cite{sociophysics} are expected to be modelled by such systems, but one is also interested in realistic  physics on complex geometries, such as for integrated nanoparticle systems \cite{Tadic05}.
Some complex networks are characterised by so-called  scale-free\index{scale-free} behavior. The degree of a node in the network is the number of links emanating from it and is denoted by $k$. With $P(k)$ symbolising the degree probability distribution (the likelihood that an arbitrary chosen node has a certain degree value), this is power-law in scale-free systems,
\begin{equation}
 P(k)\sim k^{-\lambda}.
 \label{eq18}
\end{equation}

For the critical phenomena previously described, dimensionality and length scales play crucial roles. Near the critical points itself, order-parameter fluctuations  tend to be strongly correlated, the correlation length diverges and the pair-correlation function changes from an exponential to a power-law. The FSS hypothesis (\ref{modFSS}) depends upon the ratio of length scales and the dimensionality enters the scaling relations through hyperscaling (\ref{Jo}) and (\ref{SRlog1}).

However this notion of metrics is lost when we move from a lattice substrate to a complex network.
For Euclidean lattices the coordination number is dimension dependent (it is $2d$ for a $d$-dimensional hypercube).
For networks the coordination numbers become the degrees associated with nodes and $\lambda$ plays the role of $d$.
The node degree distribution function exponent $\lambda$ in Eq.(\ref{eq18}) controls the non-homogeneity manifest in a network due to its internal structure.
This is a principal difference between the origin of logarithmic corrections on regular lattices and on networks.
It turns out that if $\lambda$ exceeds a critical value $\lambda_c=5$ the phase transition has the usual mean-field critical exponents.
As $\lambda$ decreases the node-degree distribution becomes increasingly fat-tailed and the relative amount of high-degree nodes (so-called hubs) increases.
This leads to non-trivial critical behavior.
As a result, systems with intermediate degree distributions where $\lambda_s < \lambda < \lambda_c$ have critical exponents which are generally $\lambda$-dependent. Here $\lambda_s=3$.
Decreasing $\lambda$ still further to $\lambda<\lambda_s$, the system becomes ordered at any finite temperature.
Here only an infinite temperature field is capable of destroying the order.

Just as in the regular-lattice case, where logarithmic corrections arise at $d=d_c$, so too in the network case can they emerge at a marginal value $\lambda=\lambda_c$  for a number of classical spin models.
In Ref.[\refcite{Pavo10}], the field and temperature dependencies of critical thermodynamic functions were investigated for a system with two coupled order parameters on a scale-free network. Models of this type are frequently used to describe systems with two different types of ordering.
Physical examples of such systems include ferromagnetic and antiferromagnetic, ferroelectric and ferromagnetic, structural and magnetic ordering.
A sociophysics application\cite{sociophysics} may be opinion formation where there is a coupling between the preferences for a candidate and a party in an election, for example.

A comprehensive description the system with two coupled scalar order parameters on a scale-free network was given in Ref.[\refcite{Pavo10}], where Eq.(\ref{SRlog6}) was derived. The leading exponents (there are no exponents associated with correlation functions or the correlation length) when $\lambda = 5$ are\cite{Pavo10}
\begin{equation}
\alpha=0,   \quad
\beta=1/2,  \quad
\gamma=1,   \quad
\delta=3,   \quad
\epsilon=\frac{2}{3},   \quad
\alpha_c=0, \quad
\Delta =   \frac{3}{2}.
\end{equation}
In addition, the logarithmic-correction exponents are\cite{Pavo10}
\begin{equation}
\hat{\alpha}=-1,             \quad
\hat{\beta}=-\frac{1}{2},    \quad
\hat{\gamma}=0,              \quad
\hat{\delta}=-\frac{1}{3},   \quad
\hat{\epsilon}=-\frac{2}{3},   \quad
\hat{\alpha_c}=-1,           \quad
\hat{\Delta} =-\frac{1}{2}.
\end{equation}

This completes our list of systems exhibiting multiplicative logarithmic corrections to scaling. The list is not exhaustive and the reader is invited to test the validity of the logarithmic scaling
relations in other models which exhibit these phenomena.
For example, three-dimensional anisotropic dipolar ferromagnets, in which spatial and spin degrees of freedom are linked, are experimentally accessible syetems which exhibit
logarithmic corrections\cite{Ried1995}
Systems with tricriticality also involve important logarithmic factors in three dimensions\cite{Stephen75}.

The four-dimensional diluted Ising model offers an example of a system with logarithmic corrections to leading scaling behaviour which is not power law.\cite{} The $XY$ model in two dimensions is an analagous example for a system with an infinite-order phase transition.\cite{}

\section{Conclusions}
\label{Summary}

Over the past few years, a set of  relations which link the exponents of logarithmic corrections to scaling at higher-order phase transitions has been developed. These are the logarithmic analogues of the famous scaling relations between the leading critical exponents which were developed in the 1960's and which are now pivotal in modern statistical mechanics as well as in related areas such as lattice quantum field theory. In this review, the logarithmic scaling relations are presented and tested in a number of models. In cases where there are gaps in the literature these scaling relations allow us to make predictions.

With hindsight it is perhaps surprising that these logarithmic scaling relations have not been developed earlier. As mentioned in the text, while the relations between static exponents may be derived from a suitably modified Widom hypothesis, this is not the case for the remaining exponents.
One case which may have hindered earlier attempts to develop logarithmic scaling relations is that of the Ising model in two dimensions. There, only the specific-heat dependency on the reduced temperature has a logarithm. The fact that there appears, at first sight, to be no other logarithms to relate the specific-heat logarithm to, may be a reason why scaling relations for logarithmic corrections have not been developed earlier.

Another stymying factor for an earlier development of the logarithmic scaling relations is the rather enigmatic, new, logarithmic-correction critical exponent $\hat{q}$ which characterises the FSS of the correlation length.
It was previously thought\cite{WatsonRuGa85} that the finite-size correlation length could not exceed the actual length of the system.
However, Eq.(\ref{SRlog1}) shows not just that $\hat{q}$ may exceed zero, but that it is {\emph{universal}}. This is clear because the exponents $\hat{\alpha}$ and $\hat{\nu}$, with which  $\hat{q}$ is related, are universal.
This means that (a) sensible definitions of the finite-size correlation length and (b) sensible boundary conditions must respect the universal value of $\hat{q}$. Any definition of the correlation length or any implementation of boundary conditions not respecting the universality of $\hat{q}$ can only be, in some sense, artificial.


At the upper critical dimension, where $\hat{q}$ is non-trivial, the leading critical exponents take on their mean-field values.
Therefore  they cannot be used to distinguish the universality class
(e.g., as we have seen, the leading critical exponents for the $O(n)$ model in four dimensions are all the same). Instead, the exponents of the multiplicative logarithmic corrections may be used, so that these have a similar status there to the leading exponents for $d<d_c$.

The logarithmic-correction exponents given and derived above are listed in Tables~\ref{sidewaystable1} and~\ref{sidewaystable2}.
In the Table~\ref{sidewaystable2}, bold-face symbols indicate the current gaps in the literature, which are filled by the theory presented herein and which now require independent verification.

\begin{table*}
\tbl{Leading critical exponents for the models discussed herein. }
{\begin{tabular}{lrrrrrrrrrr}
 \hline \hline    \\
 & ${\alpha}$  & ${\beta}$ & ${\gamma}$ & ${\delta}$  & ${\epsilon}$ & ${\alpha}_c$ &  ${\nu}$ &  ${\nu}_c$ & ${\Delta}$&  ${\eta}$ \\
 \hline \hline
 & & & & & & & &  & &  \\
  Pure  Ising model (2D)&   $0$ & $\frac{1}{8}$ &  $\frac{7}{4}$ & $15$ & ${0}$ & ${0}$    & $1$  & ${{0}}$ &  $0$      & $\frac{1}{4}$ \\
 & & & & & & & & & &   \\
\hline
 & & & & & & & & & &   \\
   Random-bond/site  & & &  & & & & & & &   \\
  Ising model (2D) &  $0$ & $\frac{1}{8}$ &  $\frac{7}{4}$ & $15$ & ${0}$ & ${0}$    & $1$  & ${{0}}$ &  $0$ &  $\frac{1}{4}$     \\
 & & & & & & & & & &    \\
\hline
 & & & & & & & & & &  \\
 $O(n)$ $\phi^4$ (4D) & $0$ &  $\frac{1}{2}$ & $1$ & $3$ & $\frac{2}{3}$ & $0$   & $\frac{1}{2}$ & $\frac{1}{3}$ & $\frac{3}{2}$ & $0$     \\
 & & & & & & & & & &  \\
 \hline
 & & & & & & & & & &  \\
   $4-$State Potts  & & & & & & & & & &   \\
  model (2D)  &   $\frac{2}{3}$ &  $\frac{1}{12}$ & $\frac{7}{6}$ & $ \frac{1}{15}$ & ${\frac{4}{15}}$ & ${\frac{8}{15}}$   & $\frac{2}{3}$ & ${{\frac{8}{15}}}$ & ${{\frac{5}{4}}}$ &  $\frac{1}{4}$     \\
 & & & & & & & & & &    \\
 \hline
 & & & & & & & & & &   \\
 Long-range   & & & & & & & & & &   \\
  models ($2\sigma$D)&$0$ &  $\frac{1}{2}$ & $1$ & $3$ & $\frac{2}{3}$ & $0$   & $\frac{1}{\sigma}$ & $\frac{1}{3}$ & $\frac{3}{2}$ &  $2-\sigma$     \\
 & & & & & & & & & &   \\
 \hline
 & & & & & & & & & &   \\
 $m-$ Component & & & & & & & & & &  \\
 Spin Glasses (6D)   &
$ -1$ &  $1$ & $1$ & $ 2$  & $1$ & $-\frac{1}{2}$  & $\frac{1}{2}$ & $\frac{1}{4}$ & $2$ &  $0$     \\
 & & & & & & & & & &   \\
 \hline
 & & & & & & & & & &   \\
 Percolation (6D): & $ -1$ &  $1$ & $1$ & $ 2$  & $1$ & $-\frac{1}{2}$  & $\frac{1}{2}$ & $\frac{1}{4}$ & $2$ &  $0$     \\
 & & & & & & & & & &   \\
 \hline
 & & & & & & & & & &   \\
  YL Edge (6D)                  &$ -1$ &  $1$ & $1$ & $ 2$  & $1$ & $-\frac{1}{2}$  & $\frac{1}{2}$ & $\frac{1}{4}$ & $2$ &  $0$     \\
 & & & & & & & & & &   \\
 \hline
 & & & & & & & & & &   \\
  Lattice Animals (8D)          &$ \frac{1}{2}$ &  $\frac{1}{2}$ & $\frac{1}{2}$ & $ 2$  & $\frac{1}{2}$ & $-\frac{1}{2}$  & $\frac{1}{4}$ & $\frac{1}{2}$ & $1$ &  $0$     \\
 & & & & & & & & & &   \\
 \hline
 & & & & & & & & & &   \\
  $n$-colour Ashkin-& & & & & & & & & &   \\
   Teller model (2D): & $0$ & $\frac{1}{8}$ &  $\frac{7}{4}$ & $15$ & ${0}$ & ${0}$    & $1$  & ${{0}}$ &  $0$ &  $\frac{1}{4}$ \\
 & & & & & & & & & &   \\
 \hline
 & & & & & & & & & &   \\
 Networks ($\lambda = 5$)& $0$ & $\frac{1}{2}$ & $1$ & $3$ & $\frac{2}{3}$ & $0$ & & & $\frac{3}{2}$ &   \\
 & & & & & & & & & &   \\
  \hline
\hline
\end{tabular}}
\label{sidewaystable1}
\end{table*}

\begin{sidewaystable*}
\tbl{Logarithmic-correction critical exponents for a variety of models. Bold face indicates gaps in the literature filled by the theory presented herein - i.e., bold-face values are predictions coming from the scaling relations for logarithmic corrections and remain to be verified independently. }
{\begin{tabular}{l@{\hskip -0.5cm}rrrrrrrrrrr}
 \hline \hline    \\
 & $\hat{\alpha}$  & $\hat{\beta}$ & $\hat{\gamma}$ & $\hat{\delta}$  & $\hat{\epsilon}$ & $\hat{\alpha}_c$ &  $\hat{\nu}$ &  $\hat{\nu}_c$ & $\hat{\Delta}$& $\hat{q}$ & $\hat{\eta}$ \\
 \hline \hline
 & & & & & & & &  & & & \\
  Pure  Ising model (2D)&   1 & 0 &  0 & 0 & ${0}$ & $0$    & 0  & $0$ &  0 & 0 & 0 \\
 & & & & & & & & & & &  \\
\hline
 & & & & & & & & & &  & \\
   Random-bond/site  & & &  & & & & & & & &  \\
  Ising model (2D) & 0 & $-\frac{1}{16}$ & $ \frac{7}{8}$ & $0$ & $\mathbf{-\frac{1}{2}}$ & $\mathbf{-1}$    & $\frac{1}{2}$ & ${\mathbf{0}}$ & ${\mathbf{-\frac{15}{16}}}$ & ${\mathbf{0}}$ & $0$     \\
 & & & & & & & & & & &   \\
\hline
 & & & & & & & & & & &  \\
 $O(n)$ $\phi^4$ (4D) & $\frac{4-n}{n+8}$ &  $\frac{3}{n+8}$ & $\frac{n+2}{n+8}$ & $\frac{1}{3}$ & $\mathbf{\frac{10-N}{3(N+8)}}$ & $\mathbf{\frac{4-N}{2N+8}}$   & $\frac{n+2}{2(n+8)}$ & ${\mathbf{\frac{1}{6}}}$ & $\frac{1-n}{n+8}$ & $\frac{1}{4}$ & $0$     \\
 & & & & & & & & & & &  \\
 \hline
 & & & & & & & & & & &  \\
   $4-$State Potts  & & & & & & & & & & &  \\
  model (2D)  &   $-1$ &  $-\frac{1}{8}$ & $\frac{3}{4}$ & $ -\frac{1}{15}$ & $\mathbf{-\frac{23}{30}}$ & $\mathbf{-\frac{22}{15}}$   & $\frac{1}{2}$ & ${\mathbf{\frac{1}{30}}}$ & ${\mathbf{-\frac{7}{8}}}$ & $0$ & $-\frac{1}{8}$     \\
 & & & & & & & & & &  &  \\
 \hline
 & & & & & & & & & & &  \\
 Long-range  & & & & & & & & & &  & \\
  models ($2\sigma$D)&$\frac{4-n}{n+8}$& $\frac{3}{n+8}$ & $\frac{n+2}{n+8}$ & $\frac{1}{3}$& $\mathbf{\frac{10-N}{3(N+8)}}$ & $\mathbf{\frac{4-N}{N+8}}$ &   $\frac{n+2}{\sigma(n+8)}$ & ${\mathbf{\frac{6-\sigma}{12\sigma}}}$ & ${\mathbf{\frac{1-n}{n+8}}}$& ${\mathbf{\frac{1}{2\sigma}}}$&$0$ \\
 & & & & & & & & & & &  \\
 \hline
 & & & & & & & & & & &  \\
 $m-$ Component & & & & & & & & & & &  \\
 Spin Glasses (6D)   &
$ -\frac{3m+1}{2m-1}$ &  ${\mathbf{\frac{1+m}{2(1-2m)}}}$ & $\frac{2m}{2m-1}$ & $ {\mathbf{\frac{1-3m}{4(1-2m)}}}$  & $\mathbf{\frac{1}{2}\frac{1+2m}{1-2m}}$ & $\mathbf{\frac{1}{4}\frac{7m+3}{1-2m}}$  & $\frac{5m}{6(2m-1)}$ & ${\mathbf{\frac{1}{24}\frac{5m-3}{2m-1}}}$ & ${\mathbf{\frac{1+5m}{2(1-2m)}}}$ & ${\mathbf{\frac{1}{6}}}$ & ${\mathbf{\frac{m}{3(2m-1)}}}$     \\
 & & & & & & & & & & &  \\
 \hline
 & & & & & & & & & & &  \\
 Percolation (6D): & $\frac{2}{7}$ &  $\frac{2}{7}$ & $\frac{2}{7}$ & $\frac{2}{7}$ & $\mathbf{\frac{2}{7}}$ & $\mathbf{\frac{2}{7}}$  & $\frac{5}{42}$ & ${\mathbf{\frac{5}{42}}}$ & ${\mathbf{0}}$ & ${\mathbf{\frac{1}{6}}}$ & $\frac{1}{21}$     \\
 & & & & & & & & & & &  \\
 \hline
 & & & & & & & & & & &  \\
  YL Edge (6D)                  &
  $-\frac{2}{3}$ &  ${\mathbf{0}}$ & $\frac{2}{3}$ & ${\mathbf{\frac{1}{3}}}$ & $\mathbf{0}$ & $\mathbf{-\frac{1}{3}}$   & $\frac{5}{18}$ & ${\mathbf{\frac{1}{9}}}$ & ${\mathbf{-\frac{2}{3}}} $ & ${\mathbf{\frac{1}{6}}}$ & ${\mathbf{\frac{1}{9}}}$     \\
 & & & & & & & & & & &  \\
 \hline
 & & & & & & & & & & &  \\
  Lattice Animals (8D)                  &
  $\frac{1}{3}$ &  ${\mathbf{\frac{1}{3}}}$ & $\frac{1}{3}$ & ${\mathbf{\frac{1}{3}}}$ & $\mathbf{\frac{1}{3}}$ & $\mathbf{\frac{1}{3}}$   & $\frac{1}{9}$ & ${\mathbf{\frac{1}{9}}}$ & ${\mathbf{0}} $ & ${\mathbf{\frac{1}{6}}}$ & ${\mathbf{\frac{1}{9}}}$     \\
 & & & & & & & & & & &  \\
 \hline
 & & & & & & & & & & &  \\
  $n$-colour Ashkin-& & & & & & & & & & &  \\
   Teller model (2D): & $\frac{-n}{n-2}$ & $\frac{-(n-1)}{8(n-2)}$ & $\frac{7(n-1)}{4(n-2)}$ & $0$  & $\mathbf{\frac{1-n}{n-2}}$ & $\mathbf{2 \frac{1-n}{n-2}}$   & $\frac{n-1}{n-2}$ & ${\mathbf{0}}$ & ${\mathbf{\frac{-15(n-1)}{8(n-2)}}}$  & ${\mathbf{0}}$ & $0$     \\
 & & & & & & & & & &  & \\
 \hline
 & & & & & & & & & & &  \\
 Networks &   $-1$ & $-\frac{1}{2}$ &  $0$ & $-\frac{1}{3}$ & ${\mathbf{-\frac{2}{3}}}$ & $-1$    &   &  &  $-\frac{1}{2}$ &  &  \\
 & & & & & & & & & &  & \\
  \hline
\hline
\end{tabular}}
\label{sidewaystable2}
\end{sidewaystable*}

\section*{Acknowledgements}

The author is indebted to
Bertrand Berche,
Christian von Ferber,
Reinhard Folk,
Damien Foster,
Antonio Gordillo-Guerrero,
Yurij Holovatch,
Hsiao-Ping Hsu,
Wolfhard Janke,
Des Johnston,
Christian Lang,
Juan Ruiz-Lorenzo,
Vasyl Palchykov
and
Jean-Charles Walter
with whom he has collaborated on the area of logarithmic corrections.
He is especially grateful to Yurij Holovatch for careful reading of the manuscript.
This work was supported by the 7th FP, IRSES project No. 269139 ``Dynamics and cooperative phenomena in complex physical and biological
environments'' and IRSES project No. 295302 ``Statistical physics in diverse realizations''.

\begin{appendix}[Homogeneous Functions]
\label{Ahomogeneous}

A function of a single variable $f(x)$ is said to be {\emph{homogeneous}} if multiplicative rescaling of $x$ by an amount $\lambda$ results in a multiplicative rescaling of $f(x)$ by a factor $g(\lambda)$, where $g$ is a function of $\lambda$ only:
\begin{equation}
 f(\lambda x ) = g(\lambda) f(x)
 .
 \label{homog}
\end{equation}
Thus power-laws are homogeneous functions,\index{function!homogeneous} but logarithmic functions are not. Indeed, if (\ref{homog}) holds, it turns out that both $f(x)$ and $g(x)$ have to be power laws. To see this, consider a second rescaling -- this time by a factor $\mu$. Rescaling first by $\lambda$ and then by $\mu$ leads to
\begin{equation}
 f(\mu \lambda x ) = g(\mu) f(\lambda x ) = g(\mu) g(\lambda) f(x)
 ,
\end{equation}
while rescaling by the combined factor $\mu \lambda$ yields
\begin{equation}
 f(\mu \lambda x ) = g(\mu \lambda) f(x)
 .
\end{equation}
Equating the right-hand sides of these two equations gives
\begin{equation}
 g(\mu \lambda ) = g(\mu) g(\lambda)
 .
\end{equation}
Differentiating this with respect to $\mu$ gives $\lambda g^\prime(\mu \lambda) = g^\prime (\mu) g(\lambda)$, which, after chosing $\mu = 1$ gives
\begin{equation}
 \lambda g^\prime(\lambda ) = p g(\lambda)
 ,
\end{equation}
in which $p = g^\prime(1)$.
Solving for $g$ then gives $g(\lambda) \propto \lambda^p$, so that $g(\lambda)$ is power -law.
Inserting this into Eq.(\ref{homog}) and chosing $\lambda = x^{-1}$ then gives $f(1) = x^{-p} f(x)$, or
\begin{equation}
 f(x) \propto x^p,
\end{equation}
so that $f(x)$ is also power-law.

A function of two variables $f(x,y)$ is called homogeneous if
\begin{equation}
 f(\lambda x, \lambda y) = g(\lambda) f(x,y)
 .
 \label{homog2}
\end{equation}
Following similar considerations one can easily show that $g(\lambda)$ is again power-law. A more generalised homogeneous function obeys
\begin{equation}
 f(\lambda^r x, \lambda^s y) = \lambda^p f(x,y)
 .
 \label{homog2gen}
\end{equation}
Renaming $\lambda^p \rightarrow \lambda$ one arrives at the following property for a generalised homogeneous function:\index{function!generalized homogeneous}
\begin{equation}
 f(\lambda^a x, \lambda^b y) = \lambda f(x,y)
 .
 \label{genhomo}
\end{equation}

\end{appendix}


\printindex                         
\end{document}